\documentclass[apj]{emulateapj}
\usepackage{apjfonts,longtable}
\shorttitle{The ELM Survey IV}
\shortauthors{M. Kilic et al.}

\begin{document}

\newcommand{\kms}{km s$^{-1}$}
\newcommand{\msun}{$M_{\odot}$}
\newcommand{\rsun}{$R_{\odot}$}

\title{The ELM Survey. IV. 24 White Dwarf Merger Systems$^\dagger$}

\author{Mukremin Kilic\altaffilmark{1,8},
	Warren R. Brown\altaffilmark{2},
	Carlos Allende Prieto\altaffilmark{3,4},
	S. J. Kenyon\altaffilmark{2},
        Craig O. Heinke\altaffilmark{5},
        M. A. Ag\"{u}eros\altaffilmark{6},
	and S. J. Kleinman\altaffilmark{7}
}

\altaffiltext{$\dagger$}{Based on observations obtained at the MMT Observatory, a joint facility
of the Smithsonian Institution and the University of Arizona.}
\altaffiltext{1}{Homer L. Dodge Department of Physics and Astronomy, University of Oklahoma, 440 W. Brooks St., Norman, OK, 73019, USA}
\altaffiltext{2}{Smithsonian Astrophysical Observatory, 60 Garden St., Cambridge, MA 02138, USA}
\altaffiltext{3}{Instituto de Astrof\'{\i}sica de Canarias, E-38205 La Laguna, Tenerife, Spain}
\altaffiltext{4}{Departamento de Astrof\'{\i}sica, Universidad de La Laguna, E-38206 La Laguna, Tenerife, Spain}
\altaffiltext{5}{Department of Physics, CCIS 4-183, University of Alberta, Edmonton, AB, T6G 2E1, Canada; Ingenuity New Faculty}
\altaffiltext{6}{Columbia University, Department of Astronomy, 550 West 120th Street, New York, NY 10027, USA}
\altaffiltext{7}{Gemini Observatory, 670 N. A'ohoku Place, Hilo HI 96720, USA}
\altaffiltext{8}{kilic@ou.edu}

\begin{abstract}

We present new radial velocity and X-ray observations of extremely low-mass (ELM, $\sim$0.2 \msun)
white dwarf candidates in the Sloan Digital Sky Survey (SDSS) Data Release 7 area.
We identify seven new binary systems with 1-18 h orbital periods.
Five of the systems will merge due to gravitational wave radiation within 10 Gyr, bringing
the total number of merger systems found in the ELM Survey to 24. The ELM Survey has
now quintupled the known merger white dwarf population. It has also discovered the eight
shortest period detached binary white dwarf systems currently known. We discuss the characteristics
of the merger and non-merger systems observed in the ELM Survey, including their
future evolution. About half of the systems have extreme mass ratios.
These are the progenitors of the AM Canum Venaticorum systems and supernovae .Ia.
The remaining targets will lead to the formation of extreme helium stars, subdwarfs, or
massive white dwarfs. We identify three targets that are excellent gravitational wave sources.
These should be detected by the Laser Interferometer Space Antenna (LISA)-like missions
within the first year of operation. The remaining targets are important indicators of what
the Galactic foreground may look like for gravitational wave observatories.

\end{abstract}


\section{INTRODUCTION}

Short period binary white dwarfs (WDs) are strong gravitational wave sources
and the potential progenitors of Type Ia \citep{webbink84,iben84} and
.Ia supernovae \citep{bildsten07}. The gravitational wave
radiation and the orbital decay in the shortest period systems may
be detected directly by space based missions like the Laser Interferometer Space
Antenna (LISA) and indirectly by ground based observations
\citep[see][]{brown11c}. \citet{nelemans09} lists 12 ultra-compact systems
that are guaranteed LISA sources, but predicts that LISA should detect
at least several hundred systems. 

The ELM Survey \citep{kilic10a,kilic11a,brown10,brown12} is opening a new window on short period binary
WDs and strong gravitational wave sources. After the discovery of four double WD systems
with merger times shorter than 500 Myr \citep{kilic10a}, radial velocity follow-up of the ELM WDs
found in the Hypervelocity-star survey \citep{brown06} and the SDSS Data Release 4 sample \citep{eisenstein06} led to
the discovery of 12 merger systems, tripling the number of known merging WD systems \citep{brown10,kilic11a}. 
In 2011, the ELM Survey identified the three shortest period detached binary WDs known, a 12-min orbital
period eclipsing system \citep{brown11c} and two 39-min orbital period systems \citep{kilic11b,kilic11c}.
All three systems show flux variations due to the relativistic beaming effect
and two of the three also show ellipsoidal variations due to tidal
distortions. These are the first two tidally distorted WDs ever found.
The three systems with $<$1 h orbital periods are strong gravitational wave sources.

\citet{marsh95} first demonstrated that the majority of low-mass
($\le$0.45 \msun) WDs are found in binaries, as the Galaxy is too young to
form such objects from single stars. \citet{brown11a} show that the binary
fraction of low-mass WDs is at least 70\%. This fraction goes up to 100\%
for ELM WDs with $M<$ 0.25 \msun\ \citep{kilic11a}. 
The Supernovae Progenitor Survey \citep[SPY,][]{napiwotzki01}, on the other hand,
finds a significantly lower binary fraction for typical 0.6 \msun\ WDs and only a
handful of binaries that will merge within a Hubble time \citep{napiwotzki07}. Hence, ELM WDs provide
the best opportunity to study the population of short period binary WDs.

In paper I \citep{brown10} of this series, we studied the population of ELM WDs found
in the Hypervelocity star survey. In paper II \citep{kilic11a}, we presented the SDSS Data
Release 4 systems. In paper III \citep{brown12}, we performed a targeted spectroscopic
survey of cooler ($\simeq$10,000 K) ELM WDs selected by color.
Here, we extend our survey to the SDSS Data Release 7 sample.  

Section 2 describes our target selection, radial velocity, and
X-ray observations. Section 3 presents
the orbital and physical parameters of the seven binaries that we targeted for
spectroscopic observations. The
entire population of 40 systems observed in the ELM Survey to date is presented
in Section 4 along with a discussion of the most interesting systems and trends.
Section 5 lists our conclusions and future prospects.

\newpage
\section{OBSERVATIONS}

\subsection{Optical Spectroscopy}

\citet{kleinman10} identifies 12 new ELM WD candidates in the SDSS Data Release
7 spectroscopy data, including J0106$-$1000 \citep{kilic11b}, J0923+3028
\citep{brown10}, J1518+0658 \citep{brown12}, and J0651+2844 \citep{brown11c}.
Here we focus on the remaining eight objects from this sample, plus J1056+6536 from
the SDSS Data Release 4 WD catalog \citep{eisenstein06}.
We present observations of seven targets with
reliable orbital solutions. The remaining targets need more
observations to constrain their orbital parameters and they will be
discussed in a future paper.

We used the 6.5m MMT equipped with the Blue Channel spectrograph
over several different observing runs between 2010 March and 2011 October. 
We operate the spectrograph with the 832 line mm$^{-1}$ grating in second
order, providing wavelength coverage 3650 \AA\ to 4500 \AA\ and a spectral
resolution of 1.2 \AA. All objects were observed at the parallactic angle,
and a comparison lamp exposure was obtained with every observation.
We flux-calibrate using blue spectrophotometric standards \citep{massey88}.

We measure radial velocities using the cross-correlation package
RVSAO \citep{kurtz98}. We first cross-correlate the observed spectra with a high
signal-to-noise WD template.  We then shift the observed spectra to the rest frame,
and sum them together to create a template for each object.  Finally, we
cross-correlate the spectra with the appropriate template to obtain the final
velocities for each object.  The average precision of our measurements is 20
\kms.

We compute best-fit orbital elements using the code of \citet{kenyon86},
which weights each velocity measurement by its associated error. We perform
a Monte Carlo analysis to verify the uncertainties in the orbital parameters
\citep[see][]{brown12}.

\subsection{X-ray Observations}

\subsubsection{Motivation}

The probability of neutron star companions to the majority of the objects identified
in the ELM Survey is only a few per cent. However, based on the mass function,
there are several ELM WD systems where the probability of a neutron star (which would
be spun up to a milli-second pulsar, MSP) companion is more than 10\%. Radio and X-ray
observations are essential to confirm or rule out such companions.

\citet{agueros09b} and \citet{kilic11a} discuss the importance of X-ray observations for
the identification of MSP companions to ELM WDs. Blackbody
emission from the surface of a possible pulsar companion to
the ELM WDs will be gravitationally bent, allowing observation of
$>$75\% of the neutron star surface in X rays even if the radio pulsar beam
misses our line of sight \citep{beloborodov02}. All 15 radio MSPs with precise
positions in unconfused regions of the globular cluster 47 Tuc have been clearly
detected in X rays \citep{heinke05,bogdanov06}. This result allows us to use the 47 Tuc
MSP sample (with accurate X-ray luminosities, $L_X$, due to its well-known distance)
to predict that other MSPs should have X-ray luminosities above
$L_X(0.5-6 keV)=2\times10^{30}$ erg s$^{-1}$, the minimum $L_X$ of MSPs in 47 Tuc.
Thus, deep X-ray observations can confirm or rule out the presence of MSP companions in the ELM WD
binary systems.

We obtained {\em Chandra} observations of two previously known ELM WDs,
SDSS J082212.57+275307.4 and SDSS J084910.13+044528.7 \citep{kilic10a}, to search for X-ray
emission from an MSP. Based on the mass function, there is a 15-18\% probability that
these stars have neutron star (1.4-3 \msun) companions
Neither had been previously observed in X rays since the ROSAT
All-Sky Survey \citep{voges99}, which neither detected them nor placed useful limits.

\subsubsection{Data Analysis}

We used {\em Chandra}'s ACIS-S detector in Very Faint mode to observe J082212.57+275307.4
for 2.0 ks on 2011 December 13, and J084910.13+044528.7 for 10.9 ks on 2011 March 2
(Table 1).  We used CIAO 4.3 \footnote{http://cxc.cfa.harvard.edu/ciao/} and CALDB 4.4.2
to reprocess the data including current calibrations, reducing the backgrounds using Very
Faint mode cleaning.  We constructed images in the 0.3-6 keV band, and found no X-ray
photons within the 1$\arcsec$ error circles around each source. We compute distances to
the ELM WDs using the models of Panei et al. (2007) and the SDSS photometry, and the neutral
hydrogen column density $N_H$ using the Colden tool\footnote{http://asc.harvard.edu/toolkit/colden.jsp}
\citep{dickey90}. We use PIMMS\footnote{http://asc.harvard.edu/toolkit/pimms.jsp} and the
X-ray spectrum of the faintest MSP in 47 Tuc (47 Tuc-T, 134 eV blackbody) to produce
0.5-6 keV $L_X$ upper limits, which we list in Table 1.

%TABLE1
\begin{deluxetable*}{lcccccc}
\tabletypesize{\footnotesize}
\tablewidth{0pt}
\tablecaption{X-ray Observations of ELM WDs}
\tablehead{
{Name} & {ObsID} & {Dist} & {$N_H$}     & {Expos} &  {Count rate}    & {$L_X$}\\
       &         &  (pc) & (cm$^{-2}$) &   (ks)  &  (cts s$^{-1}$) & (ergs s$^{-1}$)
}
\startdata
SDSS J082212.57+275307.4 & 12352 & 430 & $3.5\times 10^{20}$  & 2.0  &  $<2.2\times 10^{-3}$ & $<2.2\times 10^{29}$ \\
SDSS J084910.13+044528.7 & 12354 & 930 & $4.1\times 10^{20}$ & 10.9 &  $<4.2\times 10^{-4}$ & $<2.0\times 10^{29}$
\enddata
\tablecomments{99\% confidence X-ray count rate upper limits for two
  ELM WDs from {\it Chandra} X-ray observations.
  Count rate limit is in 0.3-6 keV band and $L_X$ limit is in 0.5-6 keV band.
}
\end{deluxetable*}

The 99\% confidence upper limits we calculate are an order of magnitude lower than the faintest
MSP observed in 47 Tuc, and factors of 19 and 20 fainter than the median $L_X$ of the
47 Tuc MSPs. Therefore the lack of detected X-ray emission from these two WDs is strong
evidence that their companions are not MSPs.

\section{RESULTS}

Our seven targets with optical spectroscopy data were classified as ELM WDs based on lower-resolution and signal-to-noise ratio SDSS spectra. We use
the MMT spectra to improve the model atmosphere analysis for these targets.
We perform stellar atmosphere model fits using synthetic DA WD spectra
kindly provided by D.\ Koester and an evolved version of the analysis code presented
by \citet{allende06}. We fit the flux-calibrated spectra as well as the continuum-corrected
Balmer line profiles. Using the spectral continuum provides improved constraints on
effective temperature. We compare best-fit solutions and find that the parameters differ on
average by $380 \pm 570$ K in $T_{\rm eff}$ and $0.03 \pm 0.05$ dex in $\log g$.
We take these differences as our systematic errors. We have 14-23 individual spectra for each object.
We use the fits to the individual spectra to derive a robust statistical error estimate.
\citet{eisenstein06} and \citet{kleinman10} use the SDSS spectra to derive physical parameters for the same targets.
Our parameters differ on average by $30 \pm 520$ K in $T_{\rm eff}$ and $0.17 \pm 0.08$ dex in $\log g$.

\begin{figure*}
\plottwo{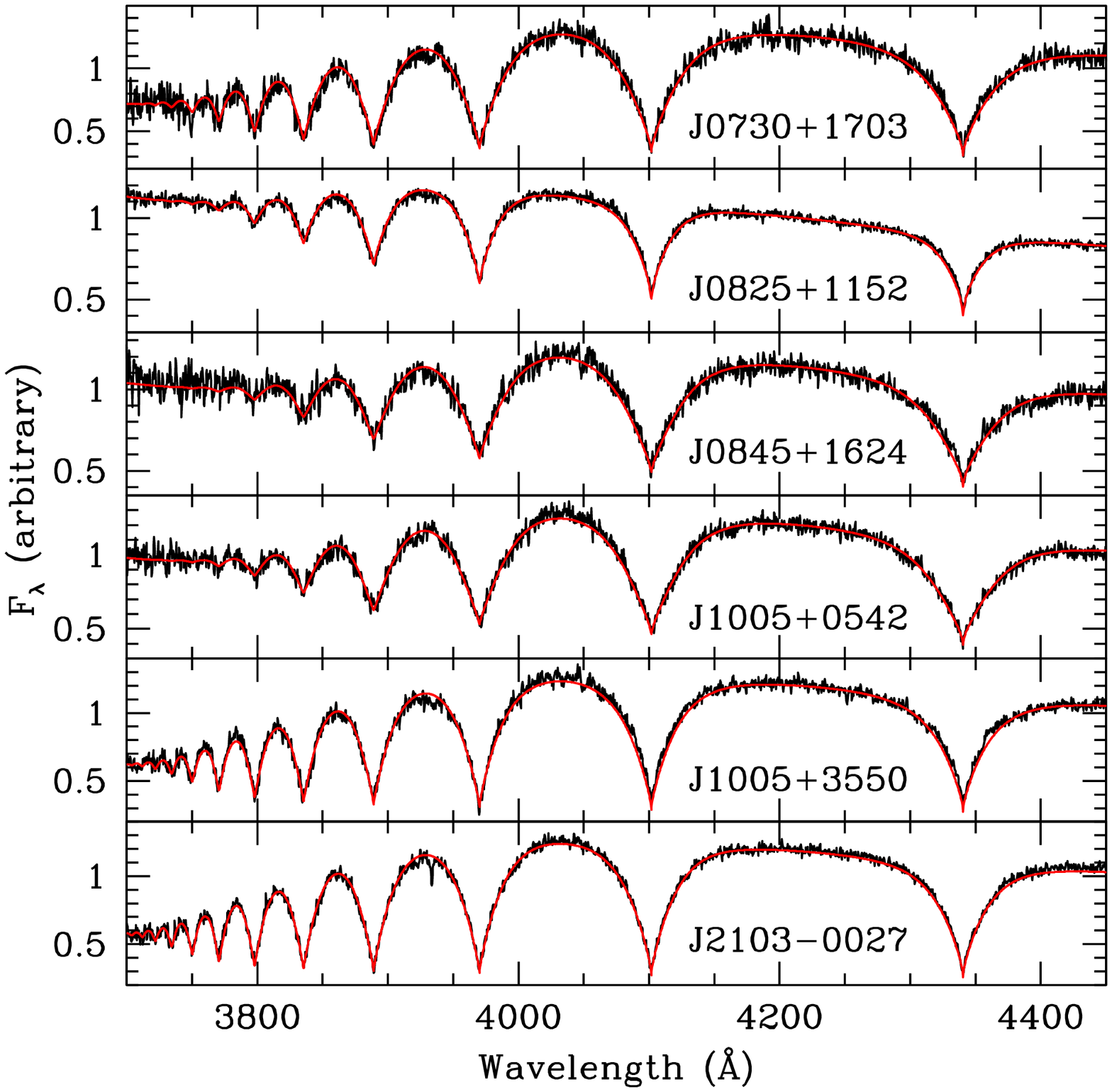}{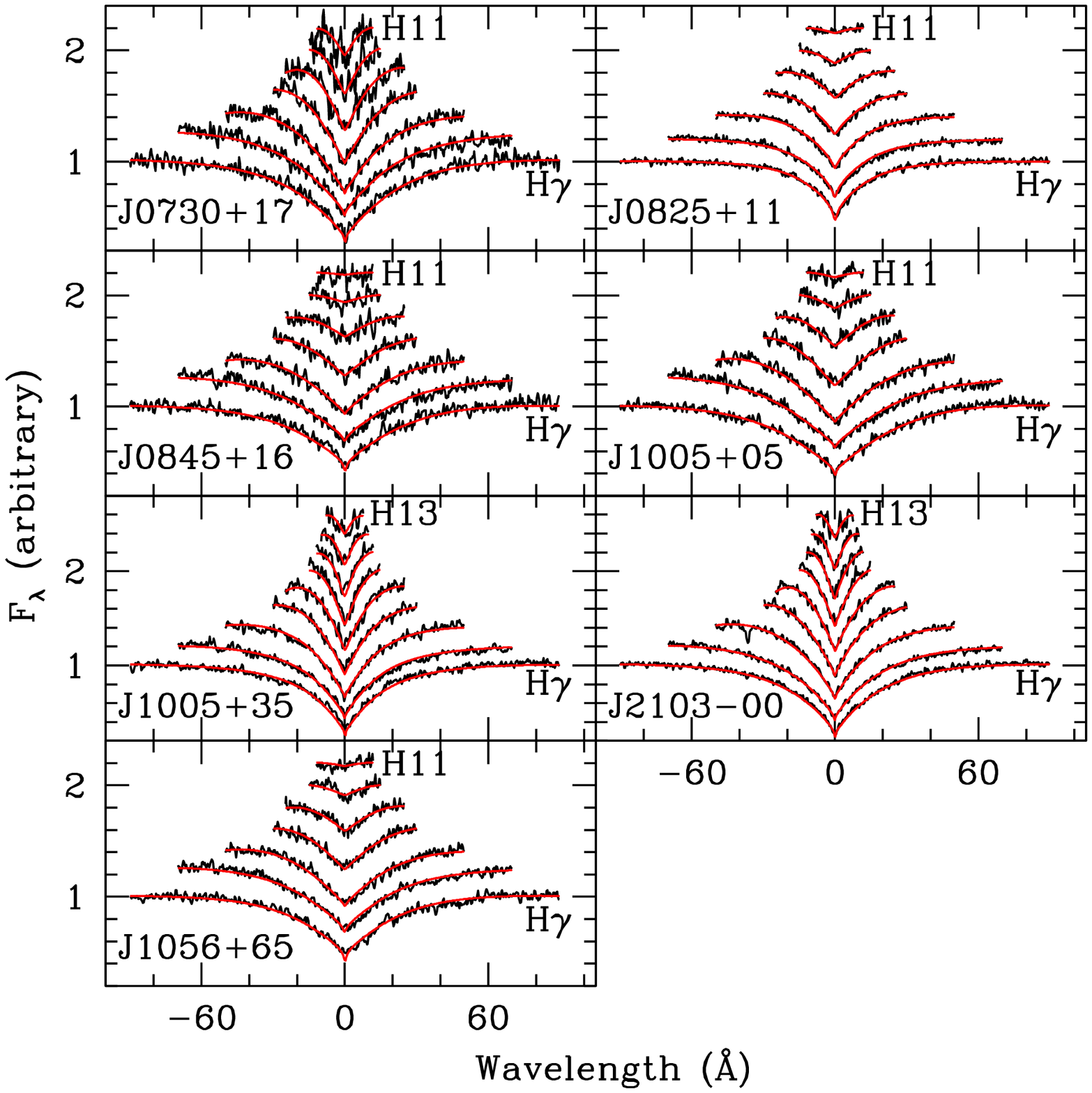}
\caption{Spectral fits (red solid lines) to the composite spectra of our targets (jagged lines, left panels)
and to the flux-normalized line profiles (right panels). The composite spectrum of J1056+6536 suffers from
flux calibration problems and is not shown in the left panels.}
\end{figure*}

Figure 1 shows the composite spectra and the best-fit models for our targets.
These models provide a good match to the observed composite spectra and the Balmer line profiles.
The parameters from both the flux-calibrated
and continuum-corrected fits are in good agreement with
the SDSS photometry in all five filters, an indication that our temperature and
surface gravity measurements are reliable. We detect flux calibration problems for
only one of our targets, J1056+6536, where the model using the continuum shape is not
a good match to the observations. We use the results from the Balmer line profile fitting
for this star. This model agrees remarkably well with the spectral energy distribution based
on the SDSS photometry.

Figure 2 compares the best-fit $T_{\rm eff}$ and $\log{g}$ measurements for our targets
against the predicted evolutionary sequences from \citet{panei07}.
Based on these tracks, our targets have $M= 0.17 - 0.40$ \msun; some of them are more massive than
predicted from the relatively noisy SDSS spectroscopy data. The physical parameters of all seven systems discussed in this
paper are presented in Table 3.
The age and distance estimates are somewhat uncertain for $M\approx0.17$ \msun\ objects, because many of them fall in the gap
between 0.17 and 0.18 \msun\ He-core WD tracks. \citet{panei07} and \citet{kilic10a}
argue that diffusion-induced hydrogen-shell flashes take place for $M>0.17$ \msun, which yield small hydrogen
envelopes. Hence, lower mass objects have massive hydrogen envelopes, larger radii,
lower surface gravities, and longer cooling times. The inconsistency between the observed parameters for
$T_{\rm eff}\sim$ 10,000 K and $\log{g}\le6$ objects and the \citet[][Figure 2]{panei07} models makes accurate WD mass
and luminosity estimates difficult for them. Fortunately, mass and luminosity change very little over the range
of effective temperature and surface gravity sampled by these WDs. We adopt $M=0.17$ \msun\ and
$M_g\simeq8$ mag for these objects.

\begin{figure}[b]
\plotone{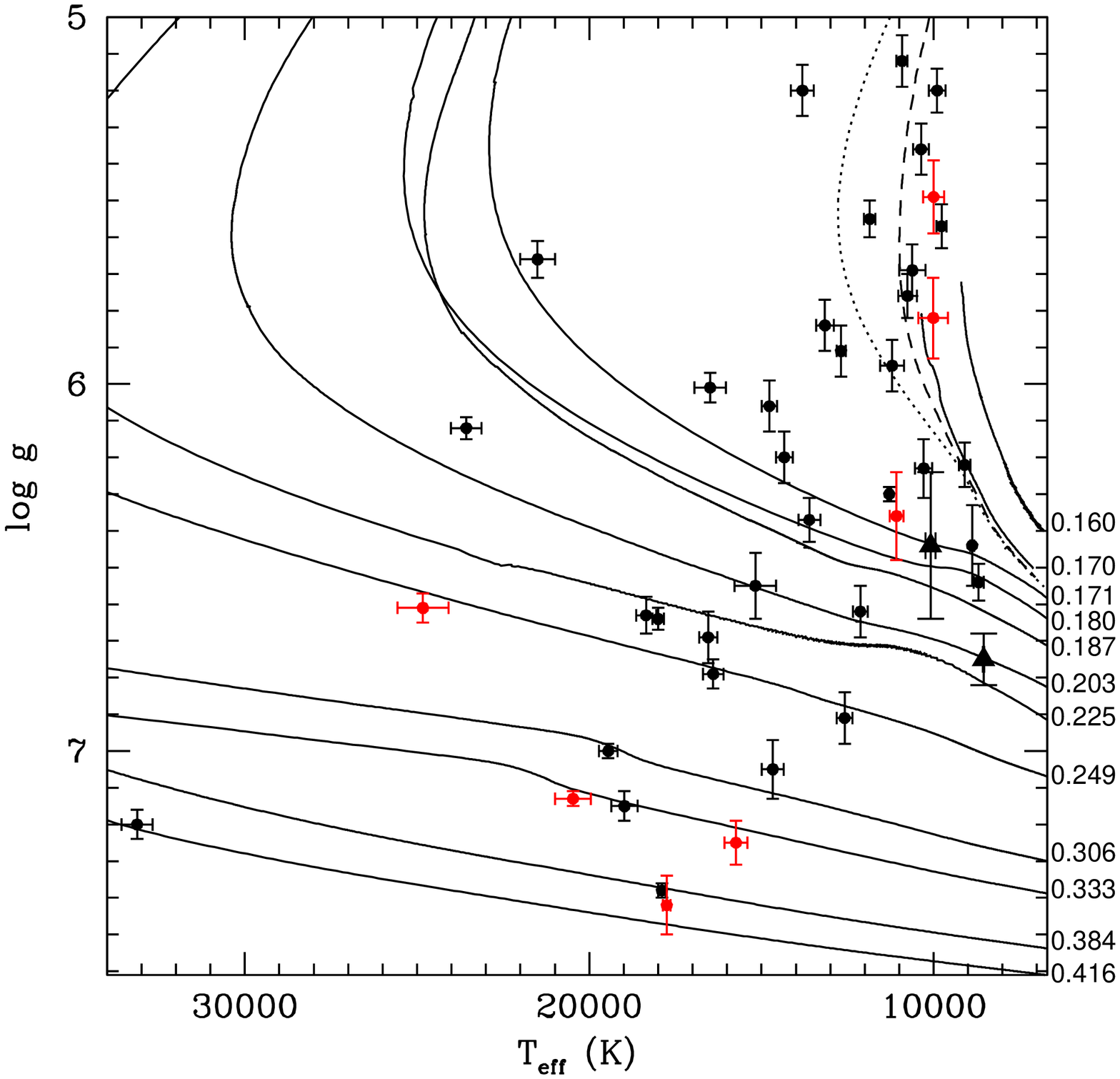}
\caption{Surface gravity versus effective temperature of the observed WDs (filled points)
in the ELM Survey, compared with predicted tracks for He WDs with 0.16--0.42 \msun\
\citep{panei07}. The dashed and dotted lines show solar metallicity and halo metallicity
(Z=0.001) models of \citet{serenelli01,serenelli02} for 0.17 $M_\odot$ WDs, respectively.
The seven new systems presented in this paper are shown as red points.
Triangles show the ELM WD companions to PSR J1012+5307 and J1911$-$5958A.}
\end{figure}

All seven targets show significant velocity variations with peak-to-peak velocity amplitudes of 120-640 \kms\ and
1-18 h orbital periods. Figure 3 shows the observed radial velocities and the best fit orbits for our targets.
We present the best-fit orbital period ($P$), semi-amplitude ($K$) of the radial velocity variations, systemic velocity
($\gamma$, which includes a small gravitational redshift term), the time of spectroscopic conjunction (the time when the primary
is closest to us), and the mass
function in Table 2. The correction for the gravitational redshift is a couple km s$^{-1}$ for a 0.17 \msun\
helium WD, comparable to the systemic velocity uncertainty. 
Based on the mass function alone, the companions would range from $M\ge0.19$ \msun\ (for J0845+1624 and J1005+3550) to
$M\ge0.71$ \msun\ objects (for J2103$-$0027). Such main-sequence companions would be detected in the
SDSS photometry and spectroscopy \citep{kroupa97}.
Hence, the companions are either more massive WDs or neutron stars (MSPs). We discuss each binary in turn.

%TABLE2
\begin{deluxetable*}{ccrrcc}
\tabletypesize{\footnotesize}
\tablecolumns{6}
\tablewidth{0pt}
\tablecaption{Orbital Parameters}
\tablehead{
\colhead{Object}&
\colhead{$P$}&
\colhead{$K$}&
\colhead{$\gamma$}&
\colhead{Spec. Conjunction}&
\colhead{Mass Function}\\
  & (days) & (\kms) & (\kms) & (HJD $-$ 2455000) & (\msun)
}
\startdata
J073032.89+170356.9   & 0.69770 $\pm$ 0.05427 & 122.8 $\pm$ 4.3 &    103.1 $\pm$ 3.4 & 276.53156 $\pm$ 0.00612 & 0.1339 $\pm$ 0.0175 \\
J082511.90+115236.4   & 0.05819 $\pm$ 0.00001 & 319.4 $\pm$ 2.7 &     32.7 $\pm$ 2.9 & 511.92318 $\pm$ 0.00146 & 0.1964 $\pm$ 0.0050 \\
J084523.03+162457.6   & 0.75599 $\pm$ 0.02164 &  62.2 $\pm$ 5.4 &     15.2 $\pm$ 4.6 & 511.69953 $\pm$ 0.01308 & 0.0188 $\pm$ 0.0049 \\
J100548.09+054204.4   & 0.30560 $\pm$ 0.00007 & 208.9 $\pm$ 6.8 &      8.6 $\pm$ 8.2 & 510.87236 $\pm$ 0.00359 & 0.2886 $\pm$ 0.0282 \\
J100554.05+355014.2   & 0.17652 $\pm$ 0.00011 & 143.0 $\pm$ 2.3 & $-$170.9 $\pm$ 1.7 & 533.04951 $\pm$ 0.00058 & 0.0535 $\pm$ 0.0026 \\
J105611.02+653631.5   & 0.04351 $\pm$ 0.00103 & 267.5 $\pm$ 7.4 &  $-$12.0 $\pm$ 4.7 & 622.85623 $\pm$ 0.00016 & 0.0863 $\pm$ 0.0074 \\
J210308.79$-$002748.9 & 0.20308 $\pm$ 0.00023 & 281.0 $\pm$ 3.2 &  $-$67.6 $\pm$ 5.3 & 384.87278 $\pm$ 0.00474 & 0.4666 $\pm$ 0.0159
\enddata
\end{deluxetable*}

\begin{figure*}
\includegraphics[width=2.5in]{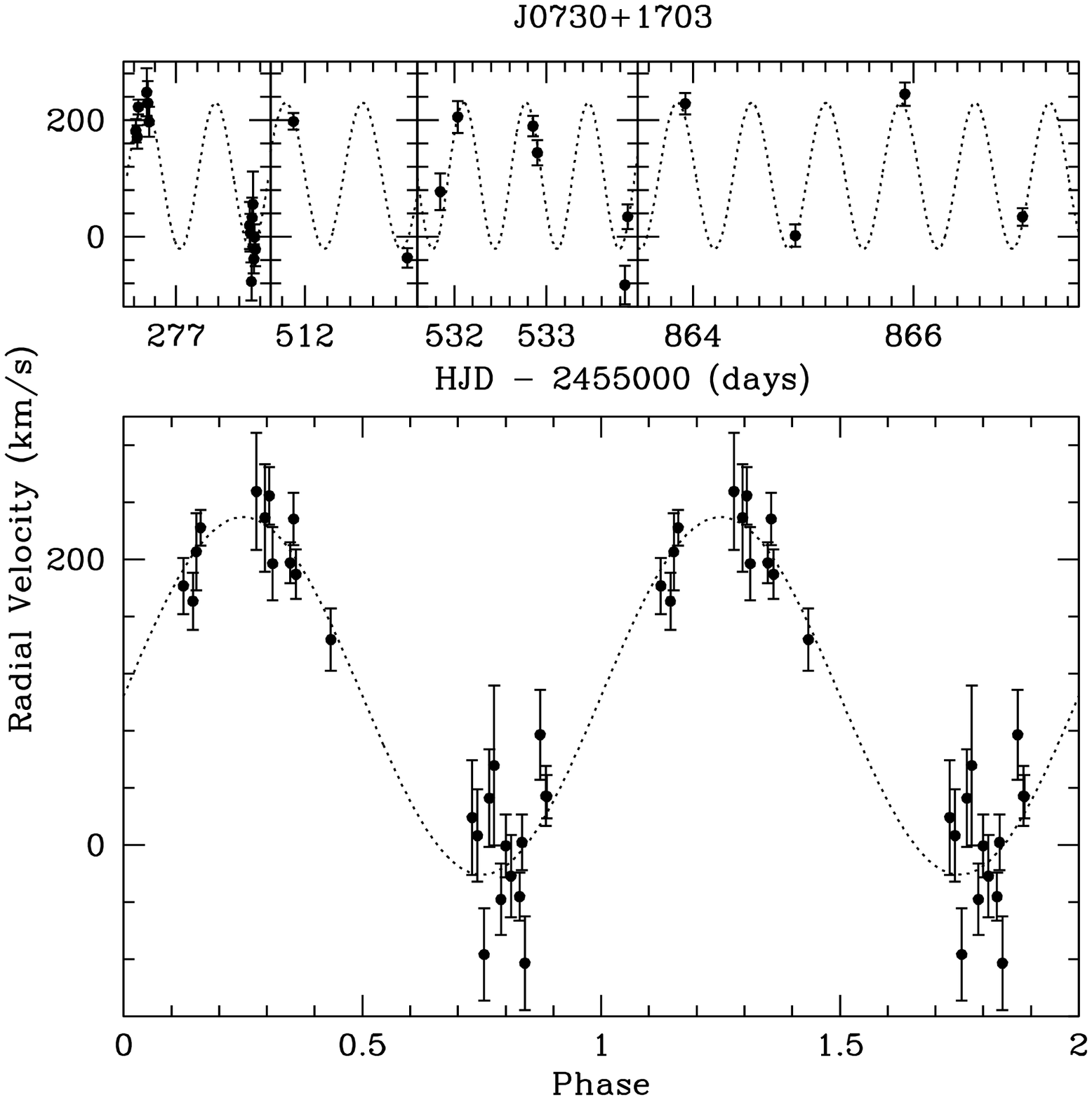}
\includegraphics[width=2.5in]{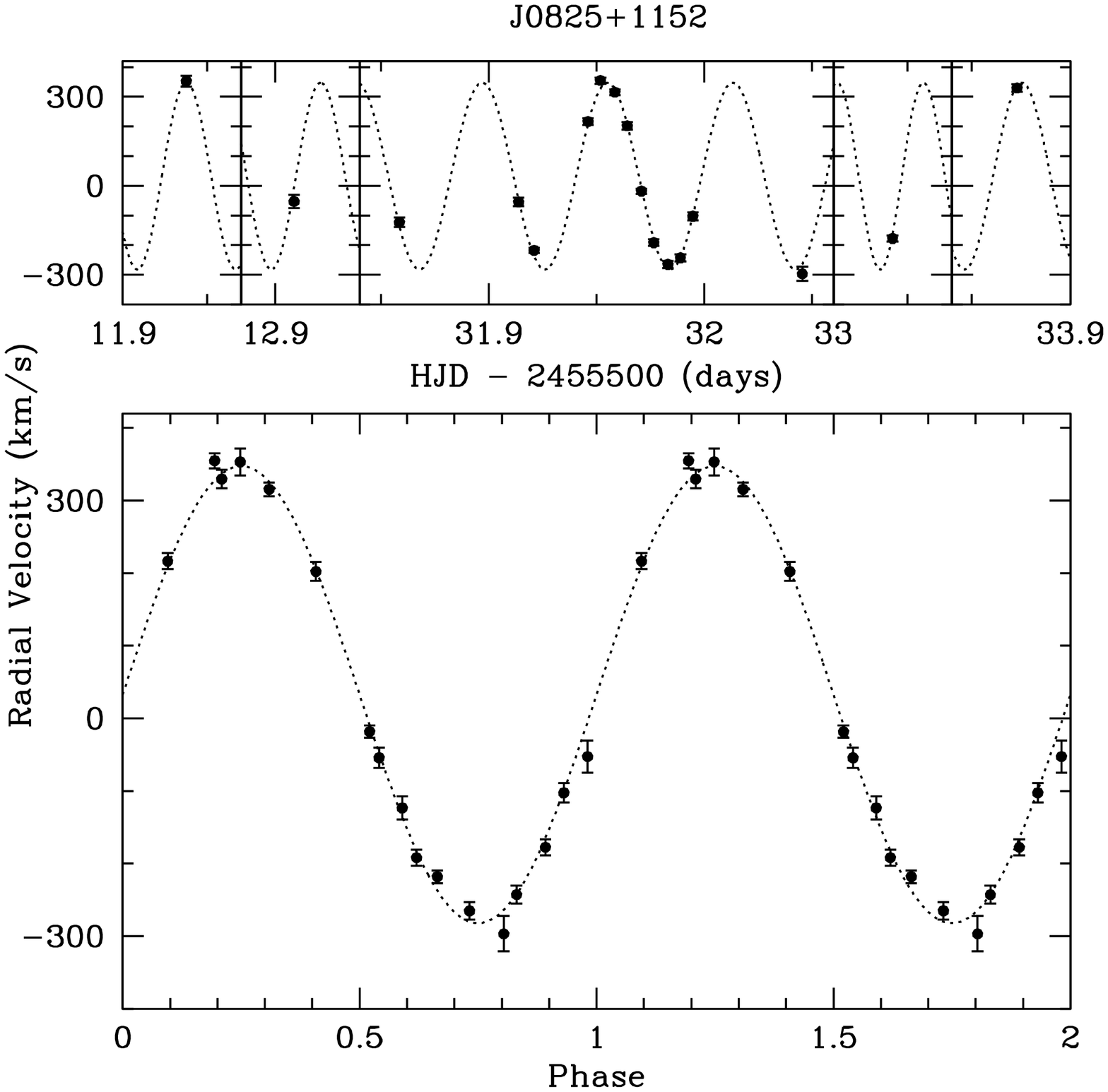}
\includegraphics[width=2.5in]{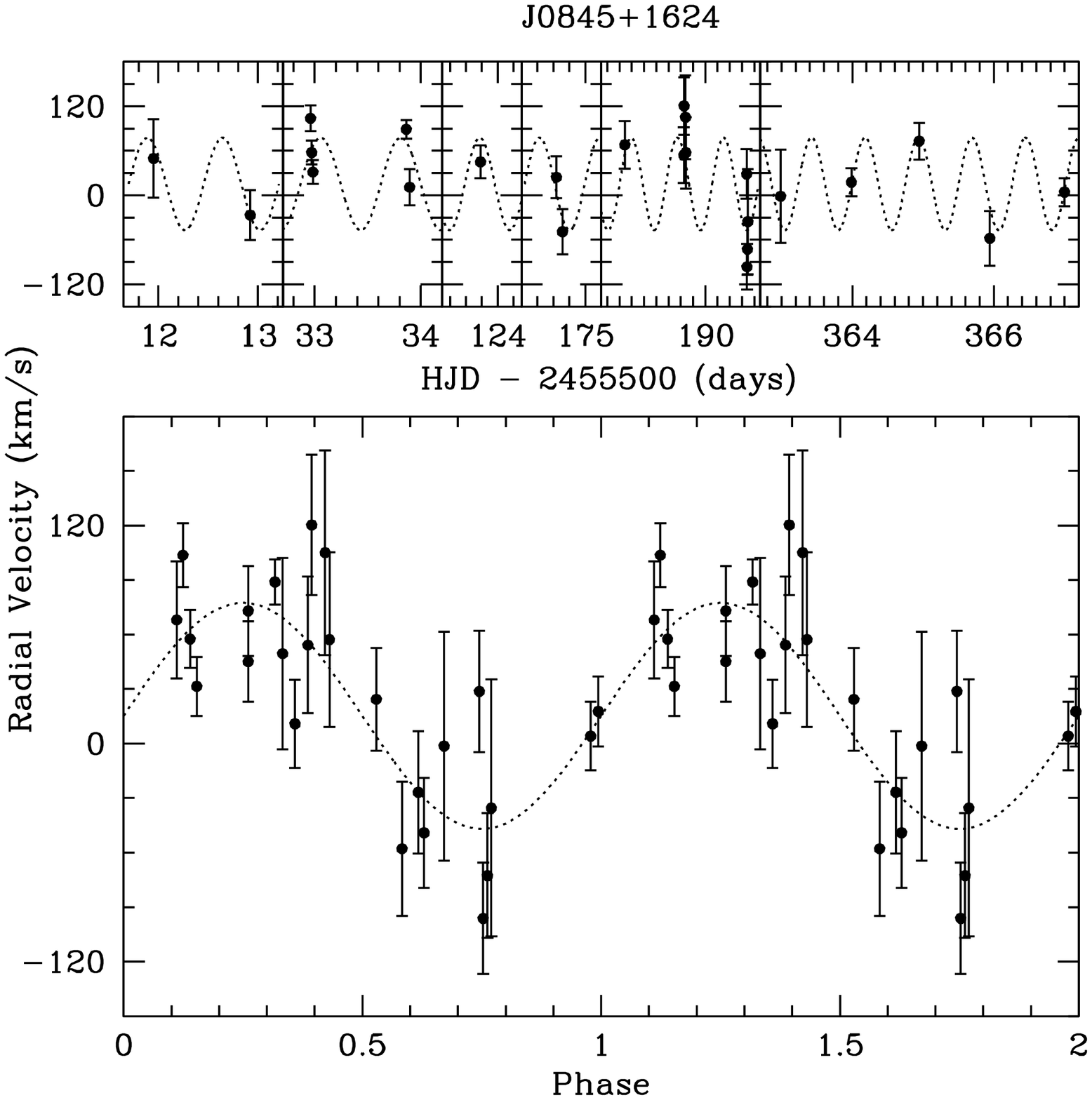}
\includegraphics[width=2.5in]{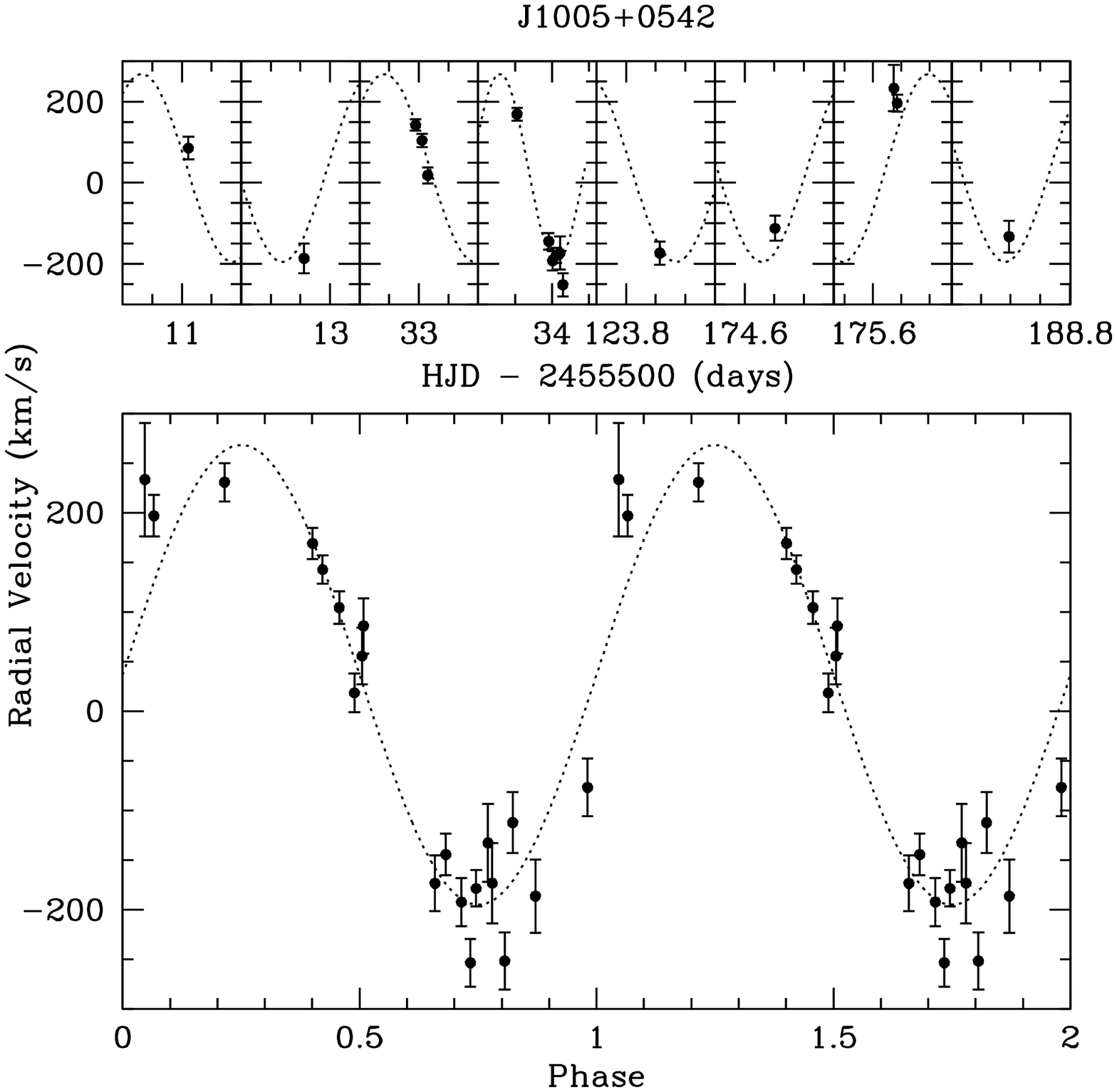}
\includegraphics[width=2.5in]{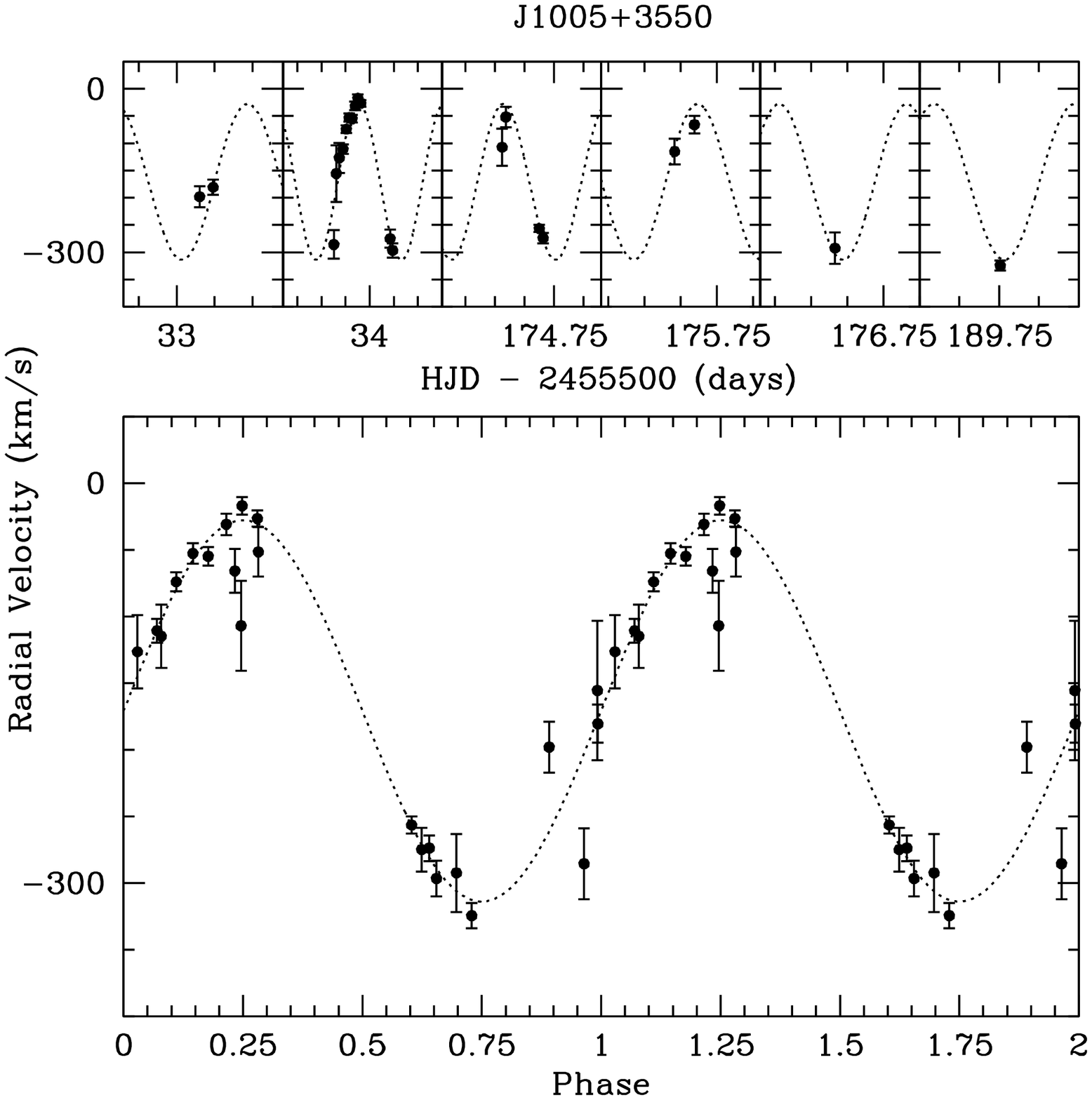}
\includegraphics[width=2.5in]{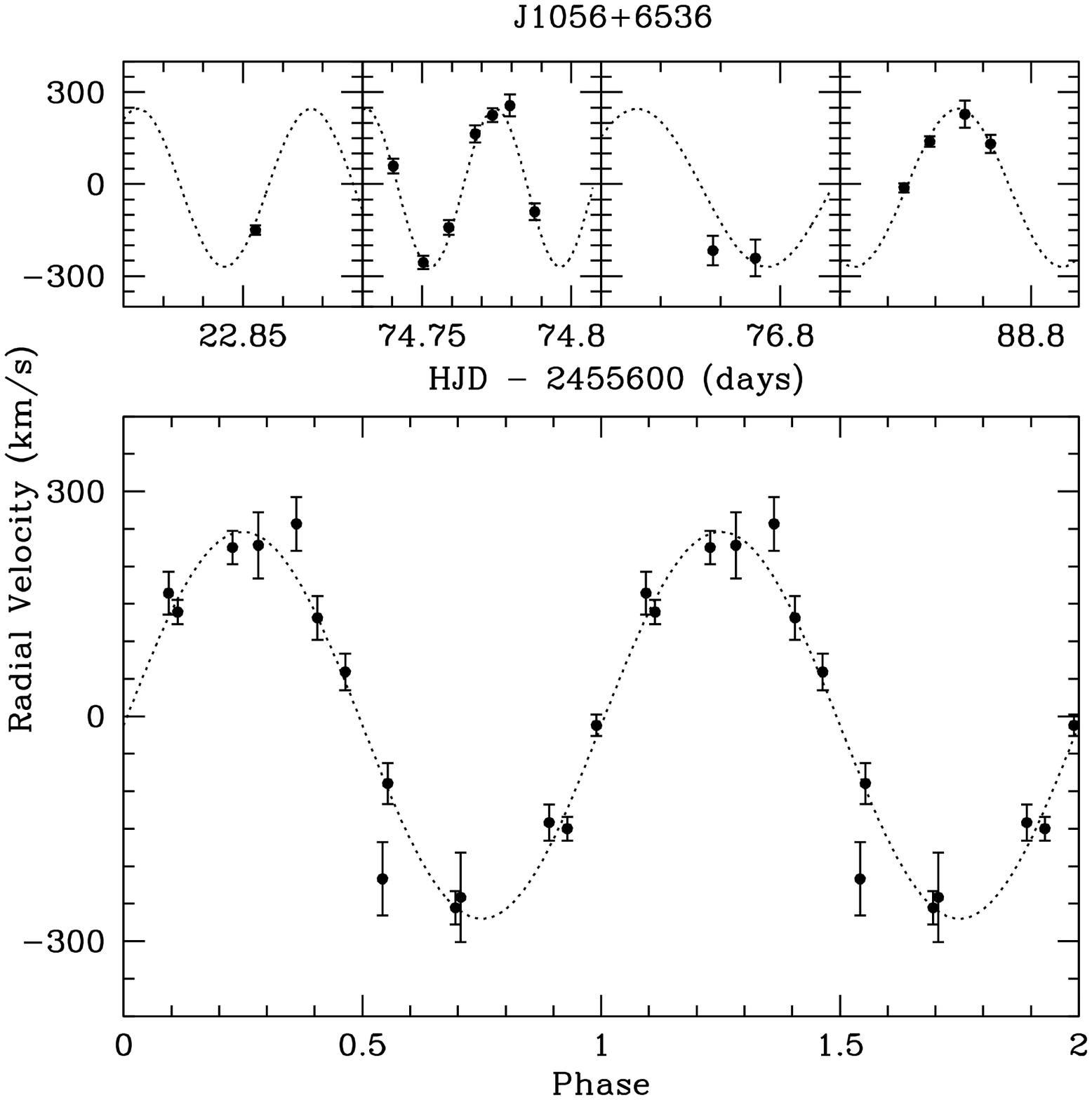}
\includegraphics[width=2.5in]{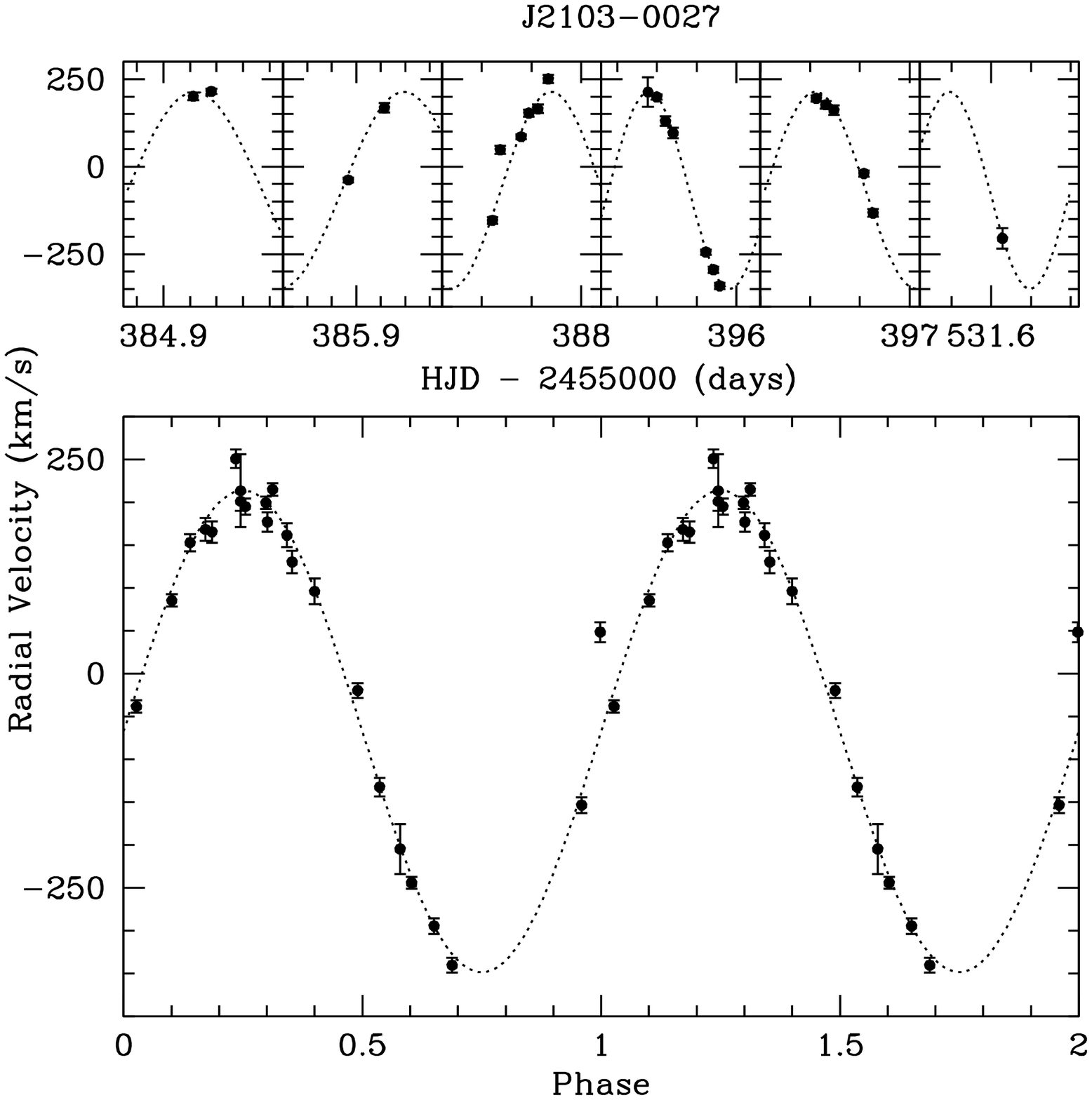}
\caption{Velocities and orbits for the 7 new ELM WD candidates. Small panels plot
the heliocentric radial velocities vs.\ observation date.  Large panels plot the
observations phased to the best-fit orbital solutions (Table 2).}
\end{figure*}

%TABLE3
\begin{deluxetable*}{ccccccccrrl}
\tabletypesize{\footnotesize}
\tablecolumns{11}
\tablewidth{0pt}
\tablecaption{Physical Parameters}
\tablehead{
\colhead{Object}&
\colhead{$g_0$}&
\colhead{$T_{\rm eff}$}&
\colhead{$\log g$}&
\colhead{Mass}&
\colhead{$M_2$}&
\colhead{$M_2(60^\circ)$}&
\colhead{$d$}&
\colhead{NS}&
\colhead{SN Ia}&
\colhead{$\tau_{\rm merge}$}\\
  & (mag) & (K) & (cm s$^{-1}$) & (\msun) & (\msun) & (\msun) & (kpc) & Prob. & Prob. & (Gyr)
}
\startdata
J0730+1703   & 19.7 & 11080 $\pm$ 200 & 6.36 $\pm$ 0.12 & 0.17 & $\ge 0.32$ & 0.41 & 1.2 & 6\% &  2\% & $\le 266$  \\
J0825+1152   & 18.6 & 24830 $\pm$ 740 & 6.61 $\pm$ 0.04 & 0.26 & $\ge 0.47$ & 0.61 & 1.6 & 9\% &  4\% & $\le 0.180$ \\
J0845+1624   & 19.7 & 17750 $\pm$ 110 & 7.42 $\pm$ 0.08 & 0.40 & $\ge 0.19$ & 0.22 & 0.9 & 2\% &  2\% & $\le 251$ \\
J1005+0542   & 19.7 & 15740 $\pm$ 330 & 7.25 $\pm$ 0.06 & 0.34 & $\ge 0.66$ & 0.86 & 1.0 & 14\% & 11\% & $\le 9.0$    \\
J1005+3550   & 18.8 & 10010 $\pm$ 430 & 5.82 $\pm$ 0.11 & 0.17 & $\ge 0.19$ & 0.24 & 1.5 & 3\% &  1\% & $\le 10.3$   \\
J1056+6536   & 19.7 & 20470 $\pm$ 520 & 7.13 $\pm$ 0.02 & 0.34 & $\ge 0.34$ & 0.43 & 1.4 & 5\% &  4\% & $\le 0.085$ \\
J2103$-$0027 & 18.2 & 10000 $\pm$ 300 & 5.49 $\pm$ 0.10 & 0.17 & $\ge 0.71$ & 0.99 & 1.1 & 17\% &  5\% & $\le 5.4$
\enddata
\end{deluxetable*}

\subsection{J0730+1703}

The ELM WD J0730+1703 has $T_{\rm eff} = 11080 \pm 200$ K and $\log{g} = 6.36 \pm 0.12$.
It is a 1.1 Gyr old 0.17 $\pm$ 0.01 \msun\ WD at a distance of 1.2 kpc.
It has a large systemic velocity of 103 \kms\, but a relatively small proper motion of
2.8 $\pm$ 4.0 mas yr$^{-1}$ in the USNO-B catalog \citep{monet03}.
Its kinematics are consistent with the disk population.

J0730+1703 has a best-fit orbital period of 16.7 h, but the current
data set allows for a significant alias at 9.9 h. The relatively small $122.8$ \kms\ radial
velocity semi-amplitude of this system implies that the companion is almost certainly low-mass,
regardless of the exact period. For the best-fit orbital period, the companion is a $M\ge0.32$ \msun\
compact object. The probability of a 1.4-3.0 \msun\ neutron
star companion is only 6\%. Assuming the mean inclination angle for a random stellar sample, $i=60\arcdeg$,
the companion is likely another low-mass WD with $M=$ 0.41 \msun. This binary will not merge within a Hubble time.

\subsection{J0825+1152}

J0825+1152 has a well-constrained orbital period of 83.79 $\pm$ 0.01 min. Because our 8 min long exposures span
10\% of its orbital phase, the observed amplitude is underestimated by a factor of 0.985. The corrected radial
velocity semi-amplitude is $K=$ 319.4 \kms.

J0825+1152 has $T_{\rm eff} = 24830 \pm 740$ K and $\log{g} = 6.61 \pm 0.04$. Based on the \citet{panei07} tracks,
it is a 40 Myr old 0.26 \msun\ WD at 1.6 kpc. Using the corrected orbital parameters, there is a 9\% probability that the
companion is a neutron star. For $i=60\arcdeg$, the most likely companion is a 0.61 \msun\ C/O WD.
The likelihood that the system contains a pair of WDs whose total mass exceeds the Chandrasekhar mass is 4\%.
The merger time due to gravitational wave radiation is less than 180 Myr.

\subsection{J0845+1624}

The SDSS spectrum of J0845+1624 is best-fit with a $T_{\rm eff} = 17430 \pm 640$ K and $\log{g} = 6.72 \pm 0.17$ model
\citep{kleinman10}, implying that it is an ELM WD. However, the SDSS spectrum is noisy for this relatively faint, $g_0=19.7$ mag (dereddened),
object. Our MMT spectrum is best explained by a model with $T_{\rm eff} = 17750 \pm 110$ K and
$\log{g} = 7.42 \pm 0.08$. J0845+1624 is a 130 Myr old 0.40 $\pm$ 0.02 \msun\ WD at 0.9 kpc.

Based on 24 different spectra, the best-fit orbital period is 18.1 h with $K=62.2 \pm 5.4$ \kms, but there are several
significant aliases (e.g. at 10.4 h). Given the relatively low-amplitude velocity variations, the orbital period is not
well constrained. J0845+1624 has a companion more massive than 0.19 \msun.
The probability of a neutron star companion is only 2\%; the companion is most likely another low-mass WD. 
This binary will not merge within a Hubble time.

\subsection{J1005+0542}

J1005+0542 was classified as an ELM WD with $T_{\rm eff} = 15190 \pm 490$ K and $\log{g} = 6.87 \pm 0.14$
based on its SDSS spectrum \citep{kleinman10}. Our higher resolution and higher signal-to-noise ratio MMT
spectrum of this $g_0=19.7$ mag object is best explained by a model with $T_{\rm eff} = 15740 \pm 330$ K and
$\log{g} = 7.25 \pm 0.06$. Hence, J1005+0542 is a 140 Myr old 0.34 \msun\ WD at 1 kpc. 

The best-fit orbital period for J1005+0542 is 7.334 $\pm$ 0.002 h. The invisible companion is a $M\ge0.66$ \msun\
compact object.
There is a 14\% probability that the companion is a neutron star. Similarly, there is an 11\%
probability that the companion is a massive WD and the combined mass of the two stars is more than 1.4 \msun.
For an average inclination angle of $i=60\arcdeg$, the companion is a 0.86 \msun\ WD.

\subsection{J1005+3550}

J1005+3550 is only 4$\arcsec$ away from a 15th mag star. To avoid contamination from this nearby source,
we kept the slit at a fixed orientation throughout the
observations. We derive $T_{\rm eff} = 10010 \pm 430$ K and $\log{g} = 5.82 \pm 0.11$ from the MMT composite
spectrum. J1005+3550 is a 0.17 \msun\ WD with an absolute magnitude of $M_g\simeq8$ and $d=1.5$ kpc. J1005+3550
has a remarkable systemic velocity of $-171$ \kms, indicating a halo origin. Unfortunately, no proper motion
measurements are available in the SDSS + USNO-B catalog \citep{munn04}.

The best-fit orbital period for J1005+3550 is 4.2 h, but there are a few significant aliases (e.g. at 3.6 h). In either case,
J1005+3550 has a merger time shorter than a Hubble time. For the best-fit orbital period of 4.2 h and $K=143$ \kms,
J1005+3550 has a relatively low-mass companion with $M\ge0.19$ \msun. For an average inclination angle of $60\arcdeg$,
the companion is a 0.24 \msun\ WD.

\subsection{J1056+6536}

J1056+6536 is the shortest period system among the seven new systems discussed in this paper. It has a well-constrained
period of 62.7 $\pm$ 1.5 min, making it the fifth shortest period detached WD system currently known.
Because our 8 min long exposures span 13\% of its orbital phase, the observed amplitude is underestimated
by a factor of 0.964. The corrected radial velocity semi-amplitude is $K=$ 267.5 \kms.

J1056+6536 was originally classified as a $T_{\rm eff}= 21910 \pm 1900$ K and $\log{g}= 7.07 \pm 0.10$ low-mass WD
by \citet{liebert04} based on an SDSS spectrum. \citet{eisenstein06} analyze the same spectrum, and find a best-fit
model with $T_{\rm eff}= 20110 \pm 630$ K and $\log{g}= 6.94 \pm 0.12$. Our higher quality MMT spectrum is best
explained by a model with $T_{\rm eff}= 20470 \pm 520$ K and $\log{g}= 7.13 \pm 0.02$. This solution is consistent
with the previous estimates within the errors. J1056+6536 is therefore a 50 Myr old 0.34 \msun\ WD at 1.4 kpc.

Based on the mass function, the companion is a $\ge0.34$ \msun\ compact object. No MSP companion
is detected in the radio data \citep{agueros09a}. Hence, the companion is almost certainly another WD. 
For an average inclination of $60\arcdeg$, it is a 0.43 \msun\ low-mass WD.
The merger time due to gravitational wave radiation is $\le85$ Myr. 

\subsection{J2103$-$0027}

J2103$-$0027 has $T_{\rm eff}= 10000 \pm 300$ K and $\log{g}= 5.49 \pm 0.10$. Like J1005+3550, we assign
$M=0.17$ \msun\ and an absolute magnitude of $M_g\simeq8$, which corresponds to a distance of 1.1 kpc.
J2103$-$0027 has a systemic velocity of $-67.6$ \kms and a proper motion of 9.3 $\pm$ 4.9 mas yr$^{-1}$ \citep{munn04}.
Its kinematics are consistent with the disk population.

J2103$-$0027 has a well constrained period of 4.874 $\pm$ 0.006 h. We observe peak-to-peak radial velocity
variations of 562 \kms. Hence, the companion is a relatively massive compact object with $M\ge0.71$ \msun.
There is a 17\% probability that the companion is a neutron star. For $i=60\arcdeg$, the companion is
a 0.99 \msun\ WD. 

J2103$-$0027 displays a Ca K line in absorption with a 0.3 \AA\ equivalent width. So far, all known ELM WDs
with $\log{g}<6$ display the Ca K line in absorption \citep{kilic07,brown10,brown12,vennes11}. The origin of Ca
is unclear, but it is most likely related to accretion from the immediate circumstellar environment
\citep[see the discussion in][]{kilic07,vennes11}.

\section{DISCUSSION}

\subsection{Seven New Binary WD Systems}

We identify seven new detached short period WD binary systems in the SDSS Data Release 7 area. All seven targets were
classified as $\log{g}<7$ WDs based on SDSS spectroscopy \citep{eisenstein06,kleinman10}. However, our MMT
data show that three of the targets have $\log{g}>7$ and $M=0.3-0.4$ \msun. Assuming the mean inclination angle
for a random stellar sample, $i=60\arcdeg$, the companions range from 0.22 \msun\ to 0.99 \msun. The probability
of neutron star companions is $\le17$\% for all systems. Hence, the companions are most likely other WDs. Radio
observations are available only for one of these targets, J1056+6536, which is also the shortest period system
in the current sample. There is no evidence of a pulsar companion in the radio data for J1056+6536. 

Five of these targets have merger times $\le10$ Gyr. The fastest merger systems are J0825+1152 and J1056+6536.
With $\tau_{\rm merge}=85$ Myr, J1056+6536 is currently the fifth fastest merger system known.

There are six targets with SDSS and/or USNO-B proper motion measurements \citep{munn04}. However, none of them show
significant proper motions, and these six systems have disk kinematics.
The only star without a proper motion measurement, J1005+3550 has a systemic velocity indicative of
halo objects, $|\gamma|=171$ \kms.

\subsection{The ELM WD Sample}

The ELM Survey has so far observed 40 low-mass WDs for radial velocity variations, discovering velocity variability
in all but four of the targets. Two objects, J0651 and NLTT 11748, are eclipsing double WD systems. No main-sequence
companions are visible in the available optical photometry and spectroscopy data for the remaining 38 targets.
There are a few cases where the probability of a neutron star companion is more than 10\%, like the intriguing system
J1741 \citep{brown12,hermes12}. Radio or X-ray data are available for 14 systems \citep[including the two sources
presented in Section 2.2,][]{vanleeuwen07,agueros09a,agueros09b,kilic11a}. No MSP companions are
detected in these data. Thus, the companions are most likely WDs. 

\subsubsection{The Period Distribution of Binary WDs}

The orbital and physical parameters for the current merger and non-merger samples
found in the ELM Survey are presented in Table 4. The orbital periods range from 12-min to 1.01 d with the median
at 4.4 h. This is significantly shorter than the median periods of 21 h and 6.7 h for more massive WDs \citep{nelemans05} and
for main sequence + He-core WD binaries \citep{zorotovic11}, respectively.  

Figure 4 shows the masses and orbital periods for our sample of ELM WDs and the previously known double WD systems
in the literature \citep{nelemans05}. Similar to the trend seen in WD + main-sequence post common-envelope binaries \citep{zorotovic11},
lower mass WDs are found in shorter period systems.
This figure demonstrates that the closest binary systems end up as ELM WDs. This is expected; shorter period systems would
start interacting earlier in their evolution compared to longer period systems and experience enhanced mass loss during the
red giant phase, hence end up as lower mass WDs. There are now three ELM WD binaries known with
$P<1$ h and nine with $P=$ 1-2 h. The ELM survey has discovered the eight shortest period double WDs currently known.

\begin{figure}
\includegraphics[width=2.6in,angle=-90]{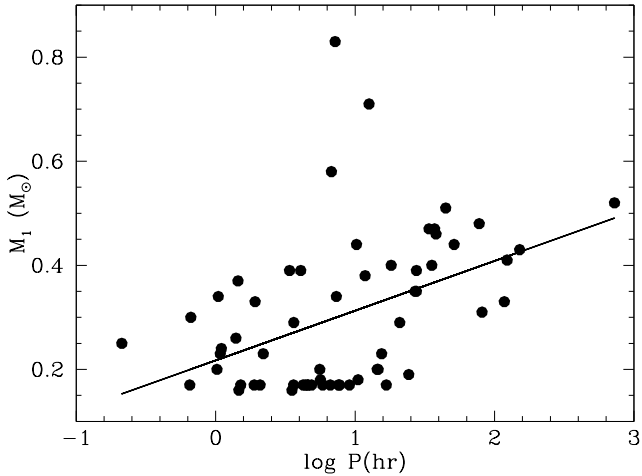}
\caption{Population of observed double WDs \citep[this paper and][]{nelemans05} as a function of orbital period and mass
of the brighter WD of the pair. The solid line is a least squares fit to the data.}
\end{figure}

We have now observed 33 WDs with $M \le 0.3$ \msun, of which 31 are velocity variable. The average velocity
semi-amplitude of these binaries is 266 \kms, whereas the upper limits for the velocity semi-amplitudes of
J0900+0234\footnote{\citet{brown12} does not rule out binarity for J0900+0234.} and J1448+1342 are 24 and
35 \kms, respectively. An average system viewed at $i \le 7.6\arcdeg$
would be consistent with the observations of these two velocity non-variable systems. For a randomly
distributed sample of orbital inclinations, there is a 0.9\% chance that $i \le 7.6\arcdeg$. Thus,
there is a 30(9)\% likelihood of finding one(two) non-variable systems in a sample of 33 stars.
It is possible that both J0900+0234 and J1448+1342 are pole-on or long-period binary systems.
Hence, the binary fraction of $M\le0.3$ \msun WDs is at least 94\% and it may be as high as 100\%.

%TABLE4
\begin{deluxetable*}{lrccrcccll}
\tabletypesize{\scriptsize}
\tablecolumns{10}
\tablewidth{0pt}
\tablecaption{Merger and non-Merger Systems in the ELM Survey}
\tablehead{
\colhead{Object}&
\colhead{$T_{\rm eff}$}&
\colhead{$\log g$}&
\colhead{$P$}&
\colhead{$K$}&
\colhead{Mass}&
\colhead{$M_2$}&
\colhead{$M_2(60^\circ)$}&
\colhead{$\tau_{\rm merge}$}&
\colhead{Ref}\\
  & (K) &  & (days) & (km s$^{-1}$) & ($M_\odot$) & ($M_\odot$) & ($M_\odot$) & (Gyr) &
}
\startdata
J0022$-$1014 & 18980  & 7.15  & 0.07989 & 145.6 & 0.33 & $\ge 0.19$ & 0.23    & $\le 0.73$  & 6\\
J0106$-$1000 & 16490  & 6.01  & 0.02715 & 395.2 & 0.17 &       0.43 & \nodata &     0.037   & 7 \\
J0112+1835   &  9690  & 5.63  & 0.14698 & 295.3 & 0.16 & $\ge 0.62$ & 0.85    & $\le 2.7$   & 1 \\
J0651+2844   & 16400  & 6.79  & 0.00885 & 657.3 & 0.25 &       0.55 & \nodata &    0.0009   & 3 \\
J0755+4906   & 13160  & 5.84  & 0.06302 & 438.0 & 0.17 & $\ge 0.81$ & 1.12    & $\le 0.22$  & 2 \\
J0818+3536   & 10620  & 5.69  & 0.18315 & 170.0 & 0.17 & $\ge 0.26$ & 0.33    & $\le 8.9$   & 2 \\
J0822+2753   & 8880   & 6.44  & 0.24400 & 271.1 & 0.17 & $\ge 0.76$ & 1.05    & $\le 8.4$   & 4 \\
J0825+1152   & 24830  & 6.61  & 0.05819 & 319.4 & 0.26 & $\ge 0.47$ & 0.61    & $\le 0.18$  & 0 \\
J0849+0445   & 10290  & 6.23  & 0.07870 & 366.9 & 0.17 & $\ge 0.64$ & 0.88    & $\le 0.47$  & 4 \\
J0923+3028   & 18350  & 6.63  & 0.04495 & 296.0 & 0.23 & $\ge 0.34$ & 0.44    & $\le 0.13$  & 2 \\
J1005+0542   & 15740  & 7.25  & 0.30560 & 208.9 & 0.34 & $\ge 0.66$ & 0.86    & $\le 9.0$   & 0 \\
J1005+3550   & 10010  & 5.82  & 0.17652 & 143.0 & 0.17 & $\ge 0.19$ & 0.24    & $\le 10.3$  & 0 \\
J1053+5200   & 15180  & 6.55  & 0.04256 & 264.0 & 0.20 & $\ge 0.26$ & 0.33    & $\le 0.16$  & 4,9 \\
J1056+6536   & 20470  & 7.13  & 0.04351 & 267.5 & 0.34 & $\ge 0.34$ & 0.43    & $\le 0.085$ & 0 \\
J1233+1602   & 10920  & 5.12  & 0.15090 & 336.0 & 0.17 & $\ge 0.86$ & 1.20    & $\le 2.1$   & 2 \\
J1234$-$0228 & 18000  & 6.64  & 0.09143 & 94.0  & 0.23 & $\ge 0.09$ & 0.11    & $\le 2.7$   & 6 \\
J1436+5010   & 16550  & 6.69  & 0.04580 & 347.4 & 0.24 & $\ge 0.46$ & 0.60    & $\le 0.10$  & 4,9 \\
J1443+1509   &  8810  & 6.32  & 0.19053 & 306.7 & 0.17 & $\ge 0.83$ & 1.15    & $\le 4.1$   & 1 \\
J1630+4233   & 14670  & 7.05  & 0.02766 & 295.9 & 0.30 & $\ge 0.30$ & 0.37    & $\le 0.031$ & 8 \\
J1741+6526   &  9790  & 5.19  & 0.06111 & 508.0 & 0.16 & $\ge 1.10$ & 1.55    & $\le 0.17$  & 1 \\
J1840+6423   &  9140  & 6.16  & 0.19130 & 272.0 & 0.17 & $\ge 0.64$ & 0.88    & $\le 5.0$   & 1 \\
J2103$-$0027 & 10000  & 5.49  & 0.20308 & 281.0 & 0.17 & $\ge 0.71$ & 0.99    & $\le 5.4$   & 0 \\
J2119$-$0018 & 10360  & 5.36  & 0.08677 & 383.0 & 0.17 & $\ge 0.75$ & 1.04    & $\le 0.54$  & 2 \\
NLTT 11748   & 8690   & 6.54  & 0.23503 & 273.4 & 0.18 &       0.76 & \nodata &      7.2    & 5,10,11 \\
\hline
J0022+0031   & 17890  & 7.38  & 0.49135 &  80.8 & 0.38 & $\ge 0.21$ & 0.26    & \nodata     & 6 \\
J0152+0749   & 10840  & 5.80  & 0.32288 & 217.0 & 0.17 & $\ge 0.57$ & 0.78    & \nodata     & 1 \\
J0730+1703   & 11080  & 6.36  & 0.69770 & 122.8 & 0.17 & $\ge 0.32$ & 0.41    & \nodata     & 0 \\
J0845+1624   & 17750  & 7.42  & 0.75599 &  62.2 & 0.40 & $\ge 0.19$ & 0.22    & \nodata     & 0 \\
J0900+0234   &  8220  & 5.78  & \nodata & $\le 24$ & 0.16 & \nodata & \nodata & \nodata     & 1 \\
J0917+4638   & 11850  & 5.55  & 0.31642 & 148.8 & 0.17 & $\ge 0.28$ & 0.36    & \nodata     & 12 \\
J1422+4352   & 12690  & 5.91  & 0.37930 & 176.0 & 0.17 & $\ge 0.41$ & 0.55    & \nodata     & 2 \\
J1439+1002   & 14340  & 6.20  & 0.43741 & 174.0 & 0.18 & $\ge 0.46$ & 0.62    & \nodata     & 2 \\
J1448+1342   & 12580  & 6.91  & \nodata & $\le 35$ & 0.25 & \nodata    & \nodata & \nodata     & 2 \\
J1512+2615   & 12130  & 6.62  & 0.59999 & 115.0 & 0.20 & $\ge 0.28$ & 0.36    & \nodata     & 2 \\
J1518+0658   &  9810  & 6.66  & 0.60935 & 172.0 & 0.20 & $\ge 0.58$ & 0.78    & \nodata     & 1 \\
J1625+3632   & 23570  & 6.12  & 0.23238 &  58.4 & 0.20 & $\ge 0.07$ & 0.08    & \nodata     & 6 \\
J1630+2712   & 11200  & 5.95  & 0.27646 & 218.0 & 0.17 & $\ge 0.52$ & 0.70    & \nodata     & 2 \\
J2252$-$0056 & 19450  & 7.00  & \nodata & $\le 25$ & 0.31 & \nodata  & \nodata & \nodata     & 2 \\
J2345$-$0102 & 33130  & 7.20  & \nodata & $\le 43$ & 0.42 & \nodata  & \nodata & \nodata     & 2 \\
LP400$-$22   & 11170  & 6.35  & 1.01016 & 119.9 & 0.19 & $\ge 0.41$ & 0.52    & \nodata     & 13,14
\enddata
\tablerefs{ (0) this paper; (1) \citet{brown12}; (2) \citet{brown10}; (3) \citet{brown11c}; (4) \citet{kilic10a};
        (5) \citet{kilic10b}; (6) \citet{kilic11a}; (7) \citet{kilic11b}; (8) \citet{kilic11c};
        (9) \citet{mullally09}; (10) \citet{steinfadt10}; (11) \citet{kawka10}; (12) \citet{kilic07};
        (13) \citet{kilic09}; (14) \citet{vennes09}}
\end{deluxetable*}

\subsubsection{Two Dozen Merger Systems}

The physical parameters and space density of double degenerate merger systems are important for
understanding the formation of R Coronae Borealis stars, single subdwarfs, AM CVn systems, supernovae Ia,
and .Ia. However, previous surveys were unsuccesful
in finding a large merger population. Figure 5 presents the total masses and periods for short period double WD systems
found in the literature \citep{nelemans05,napiwotzki07}; there were only six systems known to have short enough orbital periods
to merge within a Hubble time. We discovered 24 merger systems in a sample of 40 stars observed, a success rate of 60\%.
We have now quintupled the number of merger systems known, and we anticipate finding many more. 

\begin{figure}
\plotone{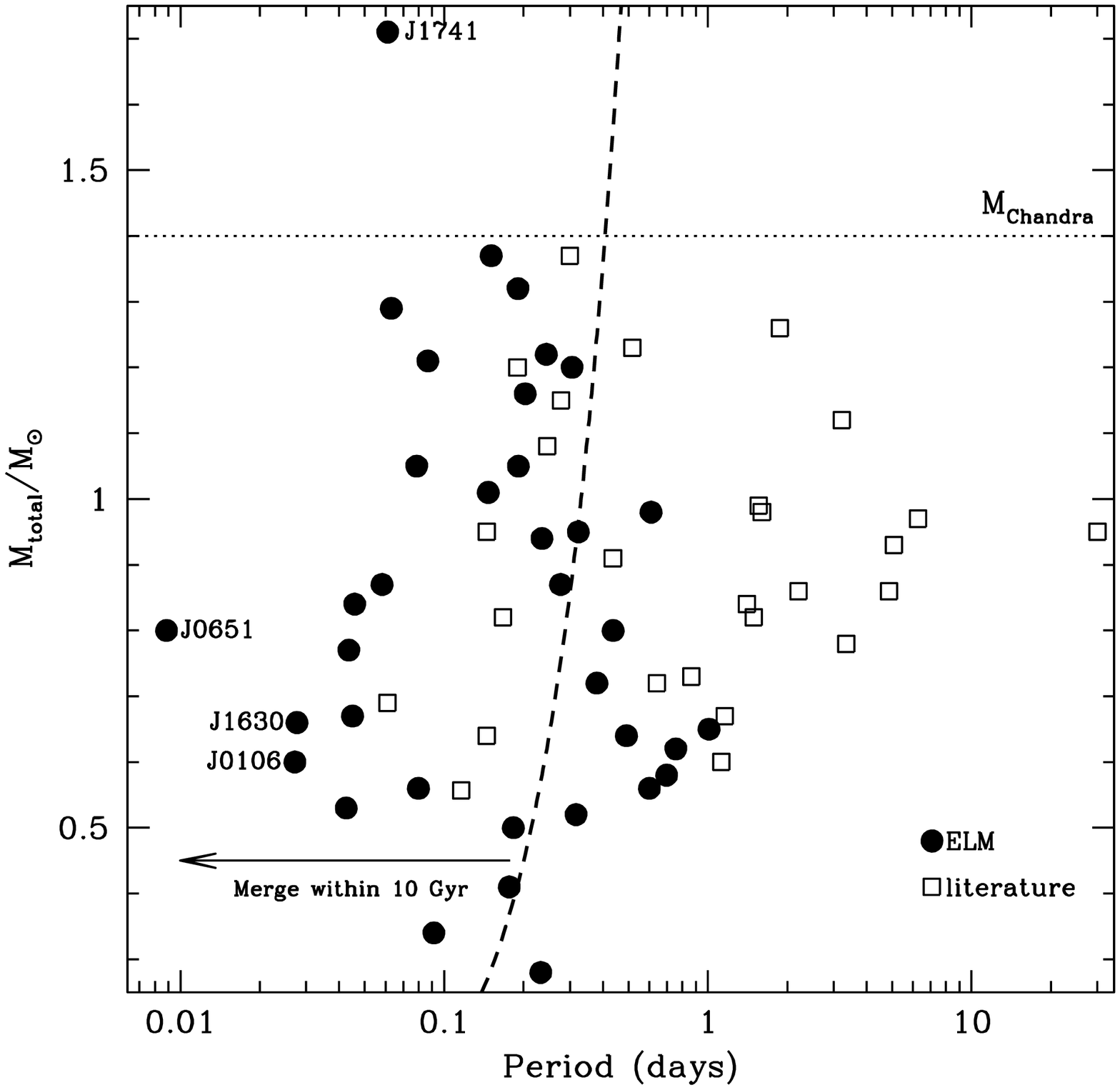}
\caption{Binary orbital period versus total system mass for the full ELM Survey and
for the previously identified double WD systems \citep{nelemans05,napiwotzki07,marsh11,parsons11,vennes11}. We
plot the total system mass assuming $i=60\arcdeg$ when the orbital inclination is unknown,
and the correct system mass when the inclination is known either from eclipses or ellipsoidal
variations. The dashed line shows the approximate threshold at
which 1:1 mass ratio systems will merge in less than 10 Gyr.}
\end{figure}

There are three more short period binary WD systems discovered in recent years; SDSS J1005+2249 \citep{parsons11}, SDSS
J1257+5428 \citep{badenes09,kulkarni10,marsh11}, and GALEX J1717+6757 \citep{vennes11}. Not surprisingly, all three systems
involve low-mass WD primaries. 

\citet{kilic11a} compare the observed period and mass distribution of double WD systems with the population synthesis
models of \citet{nelemans01a}. Based on the population synthesis calculations, \citet{kilic11a} argue that there should
not be many systems with periods less than an hour. The shortest period system known at that time was J1053+5200 with
an orbital period of 61 min. Recent discoveries of three systems with 12-39 min orbital periods indicate that $P<1$ h
detached binary WDs indeed exist. Unfortunately, the overall number distribution of the population synthesis models and
our observations cannot be directly compared due to the complicated target selection biases in the SDSS. Nevertheless,
the period distribution of our sample is informative. For example, depending on the WD cooling models used, the population synthesis calculations change significantly \citep{nelemans01a}.
The relative number distribution of short period systems in our survey can be used to constrain the population synthesis models.

Binary WDs provide important constraints on the common-envelope phase, an evolutionary stage that is difficult to study
because of its brevity. \citet{nelemans05} argue that the standard common-envelope ($\alpha$-) formalism,
equating the energy balance in the system, does not always work. Instead, they suggest that the common-envelope
evolution of close WD binaries can be reconstructed with the $\gamma$-algorithm imposing angular momentum balance.
They use the observed mass ratio distribution of double WDs ($q\sim1$) to demonstrate
that the $\gamma$ mechanism with a single value can explain all of the known systems.
Studying the prior evolution of two ELM WD binaries, \citet{kilic07} and \citet{kilic09} argue that
the same is not true for ELM WDs \citep[also see][]{woods12}. We now know that there are many ELM WDs with extreme mass ratios.
The mass distribution of our sample of ELM WDs will be extremely useful for constraining the $\gamma$ and $\alpha$
mechanisms, although with the caveat that modeling two common envelope phases for our systems clearly has large uncertainties.
\citet{demarco11} use the observed population of post-common envelope binaries including
the central stars of planetary nebulae to demonstrate that systems with small mass-ratios have higher values of
$\alpha$. Understanding the prior evolution of the ELM WDs may benefit from a similar study.

\subsection{The Future: Merger Products}

The future evolution of the ELM WD systems depend on the mass ratio of the two components. \citet{marsh04} demonstrate
that systems with extreme mass ratios of $q<<1$ will form a disk around the heavier WD, have stable mass transfer,
and turn into interacting AM CVn binaries. If the transferred mass directly impacts the accretor, it can destabilize the orbit
and lead to a merger. Figure 6 shows the observed mass distribution of the known ELM WDs and the stability criteria for different
binaries \citep[adapted from][]{marsh04,dan11}. 

\begin{figure}
\includegraphics[width=2.6in,angle=-90]{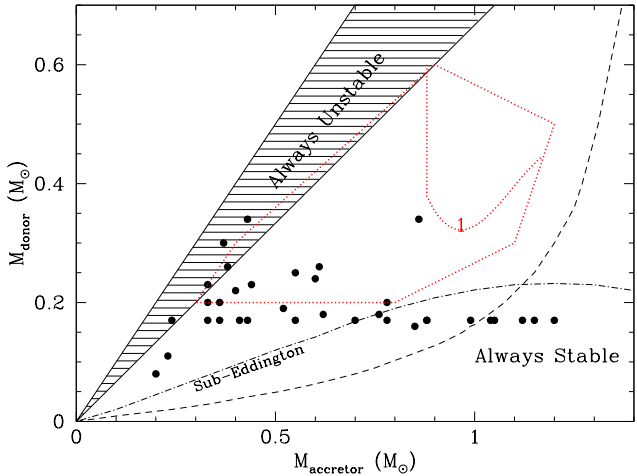}
\caption{Mass transfer stability in double WD binaries \citep{marsh04,dan11}. Binary WDs (assuming $i=60\arcdeg$ for the
systems with unknown inclinations) in the ELM Survey are shown as filled dots. Disk accretion occurs in the region below
the dashed line. These systems will evolve into stable mass transfer AM CVn. The dot-dashed line separates sub- and
super-Eddington accretion systems. Objects with mass ratios $q = 2/3 - 1$ (hatched region) will have unstable mass transfer
and merge. The area between the solid and dashed lines corresponds to either stable or unstable mass transfer depending
on the spin-orbit coupling. The dotted line marks the area studied by \citet{dan11}. The labeled contour (M. Dan 2011, private
communication) marks the region where the triple-$\alpha$ burning timescale is equal to the dynamical timescale.}
\end{figure}

The mass transfer is unstable for mass ratios $q=2/3 - 1$ and stable
for $q<<1$ (below the dashed line). The region between the solid and dashed lines corresponds to either stable or unstable
mass transfer depending on the spin-orbit coupling. The dot-dashed line separates sub- and super-Eddington accretion, with
the former leading to stable mass transfer. \citet{dan11} perform simulations for a variety of primary and secondary masses
including 0.2 \msun\ WDs (shown as dotted lines in Figure 6). All of the systems that they study, including a 0.2 + 0.8
\msun\ binary WD system, have unstable mass transfer. 

There are about a dozen ELM WDs that will have sub-Eddington accretion rates and stable mass transfer.
These systems are the progenitors of AM CVn and supernovae .Ia.
There were no known progenitors of AM CVn systems before the ELM Survey. We have now identified about a dozen potential progenitors.
\citet{brown11a} constrain the ELM WD space density using the WDs found in the magnitude-limited Hypervelocity star survey.
They find that ELM WDs contribute at least a few percent to the AM CVn population. 
The subsequent discovery of the $<$40 min orbital period
systems indicates that the ELM WD
contribution to the AM CVn population is a few times larger.
However, the current sample of ELM WDs is not a complete sample due to the SDSS target selection biases.
More accurate estimates for the ELM WD merger rate and space density have to wait until a larger magnitude-limited
survey in a well defined color-range is completed \citep{brown12}.

The remaining targets in our sample will have super-Eddington accretion rates that would lead to unstable mass transfer and merger.
There are at least five systems ($q= 2/3 - 1$) that should definitely have unstable mass transfer.
Interestingly, the eight shortest period systems will have super-Eddington accretion rates and end up as mergers.
Depending on the unknown companion mass and composition, these systems are the progenitors of extreme helium stars,
single subdwarfs, or massive WDs.

\citet{guillochon10} predict that Kelvin-Helmholtz instabilities in the accretion stream can lead to the detonation of
a surface helium layer on a CO WD and perhaps the detonation of the WD itself, a potential mechanism to initiate a Type Ia supernova from
binary WD mergers. \citet{dan11} show that the ratio of the triple-$\alpha$ burning timescale to the dynamical timescale
is of order unity for the simulations that lead to a surface detonation. This ratio is $>>1$ for our targets (Figure 6).
Hence, surface detonations through Kelvin-Helmholtz instabilities in the accretion stream are unlikely for our merging ELM WD
targets.

\subsubsection{Gravitational Wave Sources}

Double degenerate binary WDs are important gravitational wave sources. \citet{nelemans09} identifies 12 ultra-compact
binaries as LISA verification sources. There are eight AM CVn, three double degenerate,
and one ultra-compact X-ray binary sources currently known that should be detected by LISA with $\gtrsim 5 \sigma$
significance. Based on population synthesis calculations, \citet{nelemans09} estimates that at least several hundred
systems should be detectable by LISA.

Figure 7 shows the predicted gravitational wave strain amplitudes \citep{roelofs07} and frequencies of the binary systems found in the
ELM Survey. There are three binaries, J0651, J0923, and J1630, that should clearly be detected by a LISA-like mission
within the first year of operation. This is a significant addition to the LISA verification sources.
There are also three more systems that are above the 1$\sigma$ detection limit
after one year of observations, but they may be lost in the Galactic foreground of unresolved double degenerate systems.
However, it may be possible to identify these systems because we know their coordinates and physical parameters accurately
from the optical observations. The remaining 30 sources are important indicators of what the Galactic foreground may
look like for gravitational wave detectors.

\begin{figure}
\plotone{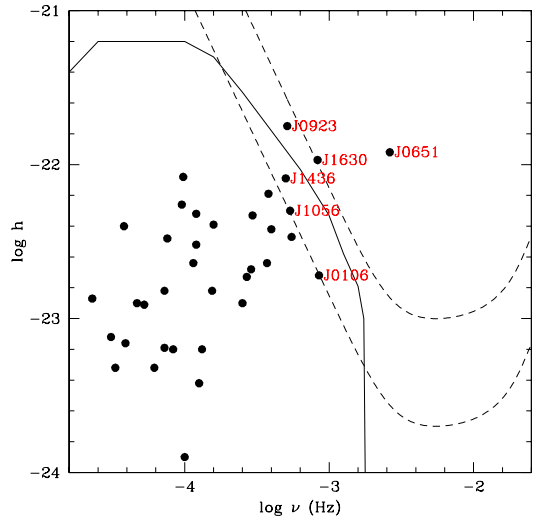}
\caption{Predicted gravitational wave strain amplitudes $h$ and frequencies $\nu$ of the binary systems found in the ELM survey.
We assume an average inclination angle of $i=60\arcdeg$, except where the inclination is known from eclipses or ellipsoidal
variations. The top and bottom dashed lines show the design sensitivities of LISA for a signal-to-noise ratio of 5 and 1,
respectively, in 1 yr of data collecting \citep{larson00}. The solid line shows the predicted Galactic foreground from
\citet{nelemans01b}. Sources above the 1$\sigma$ detection limit are labeled.}
\end{figure}

\subsubsection{Trends}

With a sample of 36 binaries discovered in the ELM Survey, we can now search for observational trends. We have
already mentioned the period differences between the ELM WD systems and the binary systems containing more massive WDs. So far,
all ELM WDs that show radial velocity variations are in $\le1$ d orbits.

Using a sample of 19 merging ELM WD systems, \citet{brown12} show that there is an absence of cool ELM WDs with short
orbital periods. Figure 8 presents periods versus temperatures for the 36 binary systems discussed here (see Table 4) and
the ELM WD companions to two MSPs, PSR J1012+5307 \citep{vankerkwijk96} and PSR J1911-5958A
\citep{bassa06}. Clearly, the shortest period systems are on average hotter than the longer period systems.
This period-temperature relation is likely because the shortest period systems merge before they cool. A 0.2 \msun\ WD
takes about 200 Myr to cool down from 20,000 K to 10,000 K \citep{panei07}, whereas the merger times for $P<0.1$ d binaries
are $\sim$100 Myr.

\begin{figure}
\plotone{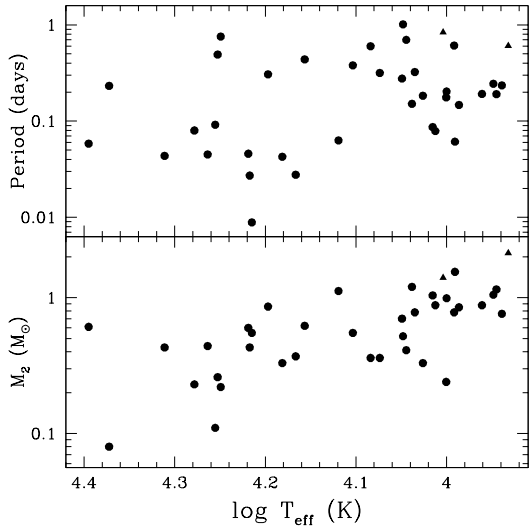}
\caption{Periods and companion masses versus temperatures for the 36 ELM WD binary systems in our sample (dots) plus the
ELM WD companions to the MSPs PSR J1012+5307 and PSR J1911-5958A (triangles).
We plot the companion masses assuming $i=60\arcdeg$ when the orbital inclination is unknown.}
\end{figure}

\citet{brown12} note that there is also a correlation between the companion mass and the ELM WD's temperature. Such a trend
is also observed for the 38 systems presented in Figure 8; hotter WDs tend to have less-massive companions.
At first glance, there is no obvious reason for such a correlation. However, our interpretation is that the shortest
period systems start interacting earlier in their evolution, go through one or two common-envelope phases relatively quickly,
and end up with lower-mass companions. Since short period systems also merge before they get a chance to cool down to
$\sim$10,000 K, there should be a deficit of cooler systems with lower mass companions. We only
include ELM WD companions to two MSPs in Figure 8, but there are
many MSPs with suspected ELM WD companions. In most cases, the companions are too faint for optical
spectroscopy. The companions to the majority of these pulsars are $\sim$10 Gyr old, hence they should
be $\sim$4000 K WDs \citep[see][]{durant12}. Hence, there are many examples of cool WDs with massive (neutron star) companions.

\section{CONCLUSIONS}

We present seven new ELM WD binary systems, including five new merger systems. The shortest period system in the new sample is
J1056+6536, with an orbital period of 62.7 min; it will merge within 85 Myr due to gravitational wave radiation. We have now
identified 24 merging WD systems in the ELM Survey, quintupling the number of double WD systems known. We present
an overview of the characteristics of this sample, including the period and mass distributions. Surface detonations
due to instabilities in the accretion stream are unlikely for these systems. Hence, we expect merging ELM WD systems
to evolve into stable mass-transfer AM CVn or unstable mass-transfer (merger) systems that would lead to the formation
of single subdwarf, extreme helium star, or massive WDs. There are about a dozen ELM WD systems with extreme mass ratios that
are potential progenitors of AM CVn systems and supernovae .Ia explosions.

We discuss the expected gravitational wave strain for our targets. There are six ELM WDs that may be detected
by a LISA-like mission with a signal-to-noise ratio of $\ge1$ after one year of observations. The remaining targets
will be part of the Galactic foreground for gravitational wave detectors at low frequencies. 
We expect that our continuing observations will lead to more merging
ELM WD discoveries, and improved constraints on the nature of
their companions and their gravitational wave signal.

\acknowledgements
We thank M. Alegria, J. McAfee, A. Milone, and J. DiMiceli for their assistance with observations at the MMT Observatory.
COH is supported by NSERC and an Ingenuity New Faculty Award. MAA gratefully acknowledges {\it Chandra} grant GO1-12019X
for support of portions of this program.

{\it Facilities:} \facility{MMT (Blue Channel Spectrograph), CXO (ACIS)}

\appendix \section{DATA TABLE}

       Table \ref{tab:dat} presents our radial velocity measurements. The table
columns include object name, heliocentric Julian date, heliocentric radial velocity,
and velocity error.\\

%\twocolumngrid
%DATA TABLE
%\begin{deluxetable}{ccr}
\begin{longtable}{ccr}
\tabletypesize{\scriptsize}
\tablecolumns{3}
\tablewidth{0pt}
\tablecaption{Radial Velocity Measurements \label{tab:dat}}
\tablehead{
\colhead{Object}&
\colhead{HJD}&
\colhead{$v_{\rm helio}$}\\
 & $-$2455000 & (\kms)
}
\startdata
J0730+1703 & 276.61626 &    181.3 $\pm$ 19.7 \\
\nodata    & 276.62942 &    170.6 $\pm$ 20.0 \\
\nodata    & 276.64002 &    222.1 $\pm$ 12.5 \\
\nodata    & 276.71921 &    247.5 $\pm$ 41.1 \\
\nodata    & 276.73155 &    229.0 $\pm$ 37.7 \\
\nodata    & 276.74215 &    196.9 $\pm$ 25.6 \\
\nodata    & 277.70070 &     19.2 $\pm$ 40.2 \\
\nodata    & 277.70785 &      6.5 $\pm$ 32.3 \\
\nodata    & 277.71731 &  $-$76.7 $\pm$ 32.5 \\
\nodata    & 277.72444 &     32.7 $\pm$ 34.2 \\
\nodata    & 277.73157 &     55.5 $\pm$ 56.1 \\
\nodata    & 277.74074 &  $-$38.2 $\pm$ 25.0 \\
\nodata    & 277.74787 &   $-$0.7 $\pm$ 22.1 \\
\nodata    & 277.75503 &  $-$21.9 $\pm$ 28.7 \\
\nodata    & 511.90070 &    197.5 $\pm$ 14.4 \\
\nodata    & 512.90032 &  $-$36.2 $\pm$ 16.8 \\
\nodata    & 531.84848 &     77.1 $\pm$ 31.4 \\
\nodata    & 532.03716 &    205.2 $\pm$ 27.1 \\
\nodata    & 532.85434 &    189.4 $\pm$ 17.3 \\
\nodata    & 532.90336 &    143.7 $\pm$ 21.9 \\
\nodata    & 533.85381 &  $-$82.8 $\pm$ 32.8 \\
\nodata    & 533.88356 &     34.2 $\pm$ 21.0 \\
\nodata    & 863.92955 &    228.2 $\pm$ 18.3 \\
\nodata    & 864.92766 &      1.7 $\pm$ 19.4 \\
\nodata    & 865.92156 &    244.4 $\pm$ 20.1 \\
\nodata    & 866.98989 &     33.8 $\pm$ 15.1 \\
\hline 
J0825+1152 & 511.93759 &    353.2 $\pm$ 18.4 \\
\nodata    & 512.91128 &  $-$52.6 $\pm$ 22.2 \\
\nodata    & 531.85847 & $-$123.4 $\pm$ 16.0 \\
\nodata    & 531.91384 &  $-$54.4 $\pm$ 13.8 \\
\nodata    & 531.92100 & $-$218.1 $\pm$  8.7 \\
\nodata    & 531.94606 &    216.5 $\pm$ 11.1 \\
\nodata    & 531.95182 &    354.8 $\pm$ 10.5 \\
\nodata    & 531.95851 &    315.3 $\pm$  9.4 \\
\nodata    & 531.96428 &    202.1 $\pm$ 12.9 \\
\nodata    & 531.97084 &  $-$18.3 $\pm$  8.6 \\
\nodata    & 531.97659 & $-$191.9 $\pm$ 10.8 \\
\nodata    & 531.98313 & $-$265.2 $\pm$ 12.0 \\
\nodata    & 531.98892 & $-$242.8 $\pm$ 12.2 \\
\nodata    & 531.99471 & $-$102.6 $\pm$ 13.5 \\
\nodata    & 532.04551 & $-$296.6 $\pm$ 24.4 \\
\nodata    & 533.03989 & $-$177.6 $\pm$ 10.9 \\
\nodata    & 533.87295 &    329.5 $\pm$ 12.5 \\
\hline 
J0845+1624 & 511.95127 &     49.5 $\pm$ 52.7 \\
\nodata    & 512.92221 &  $-$26.8 $\pm$ 33.7 \\
\nodata    & 532.96104 &    103.7 $\pm$ 17.5 \\
\nodata    & 532.97245 &     57.5 $\pm$ 16.0 \\
\nodata    & 532.98316 &     31.5 $\pm$ 16.2 \\
\nodata    & 533.86319 &     88.9 $\pm$ 12.4 \\
\nodata    & 533.89494 &     10.8 $\pm$ 24.2 \\
\nodata    & 623.78330 &     45.0 $\pm$ 22.1 \\
\nodata    & 674.63745 &     24.3 $\pm$ 28.2 \\
\nodata    & 674.71316 &  $-$49.1 $\pm$ 30.2 \\
\nodata    & 688.68475 &     68.1 $\pm$ 32.1 \\
\nodata    & 689.64873 &     54.2 $\pm$ 37.6 \\
\nodata    & 689.65468 &    120.3 $\pm$ 38.6 \\
\nodata    & 689.67658 &    105.0 $\pm$ 56.3 \\
\nodata    & 689.68299 &     57.2 $\pm$ 48.0 \\
\nodata    & 690.67644 &     28.7 $\pm$ 33.4 \\
\nodata    & 690.68216 &  $-$96.2 $\pm$ 30.6 \\
\nodata    & 690.68934 &  $-$72.6 $\pm$ 34.3 \\
\nodata    & 690.69542 &  $-$35.4 $\pm$ 70.8 \\
\nodata    & 862.98606 &   $-$1.4 $\pm$ 63.0 \\
\nodata    & 863.98615 &     17.6 $\pm$ 19.2 \\
\nodata    & 864.94448 &     72.9 $\pm$ 24.8 \\
\nodata    & 865.94358 &  $-$57.9 $\pm$ 36.9 \\
\nodata    & 866.99856 &      4.1 $\pm$ 18.9 \\
\hline
J1005+0542 & 511.01107 &     85.8 $\pm$ 27.9 \\
\nodata    & 512.95624 & $-$186.5 $\pm$ 37.0 \\
\nodata    & 532.99457 &    142.7 $\pm$ 14.1 \\
\nodata    & 533.00541 &    104.4 $\pm$ 16.4 \\
\nodata    & 533.01497 &     18.4 $\pm$ 19.7 \\
\nodata    & 533.90538 &    169.0 $\pm$ 15.6 \\
\nodata    & 533.99118 & $-$144.5 $\pm$ 20.8 \\
\nodata    & 534.00134 & $-$192.4 $\pm$ 24.2 \\
\nodata    & 534.01081 & $-$178.6 $\pm$ 18.4 \\
\nodata    & 534.02108 & $-$173.4 $\pm$ 40.5 \\
\nodata    & 534.02892 & $-$251.8 $\pm$ 28.7 \\
\nodata    & 623.85700 & $-$173.6 $\pm$ 28.1 \\
\nodata    & 674.65150 & $-$112.3 $\pm$ 30.8 \\
\nodata    & 675.63714 &    233.3 $\pm$ 57.1 \\
\nodata    & 675.64309 &    196.8 $\pm$ 21.1 \\
\nodata    & 688.69748 & $-$132.8 $\pm$ 39.4 \\
\nodata    & 863.00494 &  $-$77.0 $\pm$ 29.0 \\
\nodata    & 863.99366 &    230.6 $\pm$ 19.2 \\
\nodata    & 864.99933 &     55.5 $\pm$ 28.6 \\
\nodata    & 865.98630 & $-$253.7 $\pm$ 24.3 \\
\hline 
J1005+3550 & 533.03024 & $-$198.2 $\pm$ 19.0 \\
\nodata    & 533.04821 & $-$180.6 $\pm$ 14.3 \\
\nodata    & 533.92569 & $-$285.7 $\pm$ 26.5 \\
\nodata    & 533.93074 & $-$155.7 $\pm$ 52.4 \\
\nodata    & 533.93717 & $-$126.6 $\pm$ 27.6 \\
\nodata    & 533.94455 & $-$110.7 $\pm$  8.9 \\
\nodata    & 533.95157 &  $-$74.1 $\pm$  7.0 \\
\nodata    & 533.95768 &  $-$52.8 $\pm$  7.6 \\
\nodata    & 533.96344 &  $-$55.0 $\pm$  7.0 \\
\nodata    & 533.97009 &  $-$31.0 $\pm$  8.1 \\
\nodata    & 533.97583 &  $-$17.1 $\pm$  6.6 \\
\nodata    & 533.98159 &  $-$26.6 $\pm$  6.0 \\
\nodata    & 534.04233 & $-$275.0 $\pm$ 16.5 \\
\nodata    & 534.04776 & $-$296.7 $\pm$ 13.3 \\
\nodata    & 674.66205 & $-$107.2 $\pm$ 33.8 \\
\nodata    & 674.66829 &  $-$51.6 $\pm$ 18.6 \\
\nodata    & 674.72501 & $-$256.6 $\pm$  6.5 \\
\nodata    & 674.73147 & $-$273.9 $\pm$  9.6 \\
\nodata    & 675.69163 & $-$114.9 $\pm$ 23.8 \\
\nodata    & 675.71884 &  $-$65.9 $\pm$ 16.3 \\
\nodata    & 676.68329 & $-$292.5 $\pm$ 29.2 \\
\nodata    & 689.75147 & $-$324.6 $\pm$  9.5 \\
\hline 
J1056+6536 & 622.85315 & $-$149.9 $\pm$ 15.8 \\
\nodata    & 674.74035 &     59.2 $\pm$ 24.2 \\
\nodata    & 674.75039 & $-$255.4 $\pm$ 22.0 \\
\nodata    & 674.75893 & $-$141.8 $\pm$ 23.9 \\
\nodata    & 674.76777 &    164.4 $\pm$ 28.4 \\
\nodata    & 674.77358 &    225.2 $\pm$ 22.4 \\
\nodata    & 674.77943 &    256.7 $\pm$ 35.8 \\
\nodata    & 674.78772 &  $-$89.6 $\pm$ 27.4 \\
\nodata    & 676.78869 & $-$216.8 $\pm$ 48.7 \\
\nodata    & 676.79582 & $-$241.5 $\pm$ 59.9 \\
\nodata    & 688.77342 &  $-$12.0 $\pm$ 14.5 \\
\nodata    & 688.77879 &    139.2 $\pm$ 16.3 \\
\nodata    & 688.78617 &    228.2 $\pm$ 44.2 \\
\nodata    & 688.79154 &    131.4 $\pm$ 29.3 \\
\hline 
J2103$-$0027 & 384.92248 &    200.9 $\pm$ 10.9 \\
\nodata      & 384.93609 &    214.8 $\pm$  7.4 \\
\nodata      & 385.89359 &  $-$38.5 $\pm$  7.3 \\
\nodata      & 385.92301 &    168.1 $\pm$ 13.5 \\
\nodata      & 387.91069 & $-$153.7 $\pm$  9.2 \\
\nodata      & 387.91855 &     48.3 $\pm$ 11.7 \\
\nodata      & 387.93947 &     85.2 $\pm$  7.4 \\
\nodata      & 387.94731 &    152.7 $\pm$  9.9 \\
\nodata      & 387.95648 &    165.2 $\pm$ 12.5 \\
\nodata      & 387.96676 &    250.5 $\pm$ 10.9 \\
\nodata      & 395.88891 &    213.2 $\pm$ 42.6 \\
\nodata      & 395.89969 &    199.1 $\pm$  7.4 \\
\nodata      & 395.91083 &    130.1 $\pm$ 13.1 \\
\nodata      & 395.92039 &     95.9 $\pm$ 14.8 \\
\nodata      & 395.96147 & $-$244.4 $\pm$  6.9 \\
\nodata      & 395.97105 & $-$294.8 $\pm$  9.2 \\
\nodata      & 395.97876 & $-$340.6 $\pm$  8.7 \\
\nodata      & 396.90619 &    194.9 $\pm$  9.4 \\
\nodata      & 396.91563 &    176.7 $\pm$ 11.2 \\
\nodata      & 396.92394 &    161.4 $\pm$ 14.0 \\
\nodata      & 396.95404 &  $-$19.8 $\pm$  8.8 \\
\nodata      & 396.96331 & $-$132.6 $\pm$ 10.7 \\
\nodata      & 531.61404 & $-$204.9 $\pm$ 29.1 
\enddata
%\end{deluxetable}
\end{longtable}
\end{document}